%% file: paper_astroph_v2.tex
\documentclass[useAMS,usenatbib]{mn2e}
\usepackage{natbib}
\usepackage{graphicx}
\newcommand{\msun}{\mbox{$\,{\rm M}_\odot$}}

\title[Mass segregation and fractal substructure]{Mass segregation and fractal substructure in young massive clusters: (I) the McLuster code and method calibration}

\author[A.H.W. K\"upper, Th. Maschberger, P. Kroupa and H. Baumgardt]{Andreas
  H.W. K\"upper$^{1,2}$\thanks{E-mail: \mbox{akuepper@astro.uni-bonn.de} (AHWK); \mbox{tmasch@astro.uni-bonn.de (TM)}; \mbox{pavel@astro.uni-bonn.de} (PK);  \mbox{h.baumgardt@uq.edu.au} (HB)}, Thomas Maschberger$^{1,3}$, Pavel Kroupa$^1$\newauthor and Holger Baumgardt$^{4}$\\
$^{1}$Argelander Institut f\"ur Astronomie (AIfA), Auf dem H\"ugel 71, 53121 Bonn, Germany\\
$^{2}$European Southern Observatory, Alonso de Cordova 3107, Vitacura, Santiago, Chile\\
$^{3}$Institut de Plan\'etologie et d'Astrophysique de Grenoble (IPAG) UMR 5274, Grenoble, F-38041, France\\
$^{4}$University of Queensland, School of Mathematics and Physics, QLD 4072, Australia}

\begin{document}

\date{Accepted \ldots. Received \ldots; in original form \ldots}

\pagerange{\pageref{firstpage}--\pageref{lastpage}} \pubyear{2011}

\maketitle

\label{firstpage}

\maketitle

\begin{abstract}
By analysing models of the young massive cluster R136 in 30 Doradus, set-up using the herewith introduced and publicly made available code \textsc{McLuster}, we investigate and compare different methods for detecting and quantifying mass segregation and substructure in non-seeing limited $N$-body data. For this purpose we generate star cluster models with different degrees of mass segregation and fractal substructure and analyse them. 

We quantify mass segregation by measuring, from the projected 2d model data, the mass function slope in radial annuli, by looking for colour gradients in radial colour profiles, by measuring Allison's $\Lambda$ parameter, and by determining the local stellar surface density around each star. We find that these methods for quantifying mass segregation often produce ambiguous results. Most reliable for detecting mass segregation is the mass function slope method, whereas the colour gradient method is the least practical in an R136-like configuration. The other two methods are more sensitive to low degrees of mass segregation but are computationally much more demanding. We also discuss the effect of binaries on these measures.

Moreover, we quantify substructure by looking at the projected radial stellar density profile, by comparing projected azimuthal stellar density profiles, and by determining Cartwright \& Whitworth's $Q$ parameter. We find that only high degrees of substructure affect the projected radial density profile, whereas the projected azimuthal density profile is very sensitive to substructure. The $Q$ parameter is also sensitive to substructure but its absolute value shows a dependence on the radial density gradient of the cluster and is strongly influenced by binaries.

Thus, in terms of applicability and comparability for large sets of $N$-body data, the mass function slope method and the azimuthal density profile method seem to be the best choices for quantifying the degree of mass segregation and substructure, respectively. The other methods are computationally too demanding to be practically feasible for large data sets.

\end{abstract}

\begin{keywords}
galaxies: star clusters: individual: R136 ---  galaxies: star formation --- methods: data analysis --- Magellanic Clouds\end{keywords}

\section{Introduction}\label{Sec:Introduction}
Understanding the process of star cluster formation is vital for astrophysics since most, if not all stars are born in a clustered mode \citep{Lada03}. In the commonly accepted picture, star clusters form in three stages: first, a cold molecular cloud collapses and forms stars along filaments throughout this collapse. Secondly, the massive O- and B-stars start radiating off the residual gas until only a more or less bound ensemble of stars is left. Subsequently, the newly formed star cluster evolves dynamically until total dissolution (e.g. \citealt{Kroupa01b, Portegies10}).

In this picture, the survival of star clusters throughout the birth process and the duration of the subsequent dynamical dissolution depends crucially on the star formation efficiency, i.e. how much of the cold gas gets transformed into stars. Moreover, it depends on the structure of the gas cloud and the distribution of the forming stars. That is, for a more massive ensemble of stars the gas expulsion process will be more violent, whereas for low-mass configurations gas expulsion will happen more adiabatically \citep{Geyer01, Kroupa02b, Goodwin06, Bastian06, Baumgardt07, Baumgardt08b}. A further influence on the survival rate is given by the degree of mass segregation, i.e. when the most massive stars of an ensemble are located deeper within the forming cluster, they will have a more destructive influence on the subsequent evolution of the star cluster \citep{Vesperini09}. A yet open question in this respect is whether or not star clusters form with primordial mass segregation, or if the initial mass function (IMF) of stars is the same throughout the whole star forming complex. 

Observations of young clusters indeed suggest the existence of primordial mass segregation (e.g. \citealt{Stolte02, Bontemps10b}). But such observed mass segregation is not necessarily due to variations of the IMF, as mass segregation can also develop quickly during the first few 100,000 years of cluster formation through dynamical relaxation. The timescale of this process is proportional to the relaxation time of the configuration and the mass ratio of the forming stars \citep{Spitzer87}. Moreover, recent investigations show that initially substructured configurations can develop mass segregation on significantly shorter timescales \citep{McMillan07, Allison09, Moeckel09, Allison10, Yu11}. In this picture, (fractal) substructures, which may have much shorter relaxation times, can segregate before they merge to form the final star cluster. Hence, substructure in young star clusters is not only a sign of a system not being virialised, but may also play a vital role in the process of star cluster formation. Thus, two major aspects of young star clusters have to be investigated in detail at the different stages of star cluster formation: the degree of mass segregation and the degree of substructure.

These questions may be addressed by means of collisional $N$-body computations. But such numerical investigations also require a choice of initial conditions. The question therefore remains what configuration star clusters have at the stage where dynamical investigations can set in. From hydrodynamical computations of collapsing gas clouds it appears that star formation may be happening in a hierarchical, fractal fashion \citep{Klessen01, Bonnell03, Bonnell06}. During the subsequent dynamical evolution this substructure is erased and indeed a significant degree of mass segregation is established \citep{Maschberger10}. But hydrodynamical computations only reach gas cloud masses of a few $10^3\msun$, thus they do not shed any light on star formation in starburst regions or on the formation of globular clusters, nor can they account for the self-regulation induced by stellar feedback, yet.

Observations of young embedded star clusters show a similar picture like hydrodynamical computations. For example, \citet{Lada03}, \citet{Teixeira06}, \citet{Allen07} as well as \citet{Gennaro11} find many young clusters to show substructure and to be asymmetric. Moreover, \citet{Gutermuth05, Gutermuth08} find young embedded clusters in near-IR data to be azimuthally asymmetric with a high degree of substructure. The star formation sites appear filamentary and elongated over scales of several parsec. Such observations suggest that young clusters expand, get more symmetric and lose substructure with ongoing gas removal \citep{Gutermuth08, Bontemps10a}. The nearby, more evolved Orion Nebula Cluster, for example, shows a high degree of mass segregation which appears to be inconsistent with its current relaxation time \citep{Hillenbrand98, Kroupa02a}. In this picture, this may indicate that the ONC also formed with a high degree of substructure. 

But does this picture also apply to starbursts, and to the formation of globular cluster-like objects? Massive star formation sites with stellar masses above $10^4\msun$ in the Milky Way like NGC 3603, Westerlund 1 and the Arches cluster, are rare and mostly heavily obscured by interstellar dust. Nevertheless, mass segregation as well as high degrees of substructure have been reported for those objects (e.g. \citealt{Stolte02, Stolte06, Brandner08, Gennaro11}).  

The most massive star formation site in the Local Group is the 30 Doradus complex in the Large Magellanic Cloud. It is the only nearby starburst region, which makes it the ``Rosetta Stone'' for understanding events such as the formation of globular clusters \citep{Walborn91}. This paper therefore addresses this star formation site and aims at investigating mass segregation and substructure in this complex. For this purpose, we create models of the young massive cluster R136 which is forming in the 30 Doradus complex (Sec.~\ref{Sec:R136} \& \ref{Sec:Models}). We then use various methods from the literature (Sec.~\ref{Sec:Methods}) to detect and quantify mass segregation and substructure (Sec.~\ref{Sec:Results}). Our aim is to test and calibrate these methods in order to be able to apply them to $N$-body computations which will be presented in a follow-up investigation, and to discuss their applicability to observational data. Furthermore, Appendix A contains a manual on our new star-cluster initialisation code \textsc{McLuster}.

\section{R136}\label{Sec:R136}
The young massive cluster R136 is at the heart of the 30 Doradus (30 Dor) star forming region in the Large Magellanic Cloud (LMC). 30 Dor is known as the largest HII region in the Local Group with more than $8\times 10^5\msun$ of ionised gas within a radius of about 100 pc \citep{Kennicutt84, Malumuth94}. The densest region of 30 Dor is the central star cluster NGC 2070 with a half-light radius of about 22 pc. R136, the centre of this cluster, has always posed a challenge to high-resolution observations. This object is crucial for our understanding of star formation, since there is no other comparable starburst site in the local Universe. But due to its distance, its high central brightness and its substantially varying extinction, R136 is hard to access observationally (e.g. \citealt{Brandl96}).

R136 was found to be about 3 Myr old, but its constituent stars show some age spread \citep{Bosch01, Andersen09}. Due to  the absence of red supergiants and due to the presence of several Wolf-Rayet stars, the age of the oldest population of stars within R136 can be limited to 3--5 Myr \citep{Brandl96}. Some O stars in the centre of R136 may be less than 2 Myr old, though, indicating a complex star formation history \citep{Massey98}. The metallicity of the whole 30 Dor complex was found to be about half solar value, i.e. $Z \simeq 0.01$ \citep{Lebouteiller08}.

Mass estimates for the stellar component of R136/NGC2070 range from about $1.4\times 10^4\msun$ \citep{Malumuth94}, estimated from HST $UBV$ photometry, to about $4.5\times10^5\msun$ \citep{Bosch09}, estimated from multi-epoch stellar velocity dispersion data. Recent high precision photometry yields masses between $\sim5.5\times10^4\msun$ \citep{Crowther10}, and $> 10^5\msun$ \citep{Andersen09}.  

R136 is an important test bed for our understanding of the initial mass function of stars, and thus has been the subject of several mass-function investigations. Especially the mass function at the high-mass end and a probable upper limit of stellar masses have been primary targets of such investigations. Down to about $1.1\msun$ the mass function of R136 was found to be in agreement with a Salpeter slope of 2.35 \citep{Hunter95, Andersen09}, with some of these investigations reporting small radial dependencies of the mass function \citep{Brandl96, Selman99, Sirianni00}. 

The masses of the brightest objects in R136 are very uncertain. Due to resolution limits, R136 was once believed to be a single supermassive star of more than $1,000\msun$. Only with Speckle interferometry and the superior resolution of the HST it was later found to be a very dense star cluster (see e.g. \citealt{Weigelt91}). Recently, \citet{Crowther10} identified four stars in the centre of R136 to have masses between 165--$320\msun$, again challenging the commonly believed upper stellar mass limit of about $150\msun$ \citep{Weidner04}.

R136 most probably has a high binary fraction. \citet{Selman99} and \citet{Bosch09} find that at least all O- and B-stars are in binaries, even though \citet{Crowther10} find that the four most massive stars are most likely single stars. But the observed velocity dispersion seems to be dominated by binary motion \citep{Bosch01, Bosch09}. 

Like many young clusters in the LMC, R136 follows a shallow power-law density profile without any visible truncation at large radii, as specified by \citet{Elson87}, 
\begin{equation}\label{eq:EFF}
\rho (R) = \rho_0 \left( 1+R^2/a^2\right)^{(-\gamma/2)},
\end{equation}
where $R$ is the projected radius, $\rho_0$ is the central density, $a$ is a scale radius and $\gamma$ is the power-law slope. For R136's number density profile, $\gamma$ was found to be about 1.85 for massive stars \citep{Hunter95}. Its surface brightness profile shows a variety of power-law slopes, depending on the instrument and filter which was used. The $\gamma$ values lie between 1.7 in F336W \citep{Campbell92} and about 2 in F555W \citep{Selman99}. \citet{Andersen09} found values of $\gamma = 1.54$ and $a = 0.025$ pc in F160W, whereas according to \citet{Mackey03} $\gamma = 2.43$ for F555W and F814W. \citet{Campbell10} measured $\gamma = 1.8$ in optical data (V and I), but $\gamma = 1.6$ in near-infrared (H and K) images. \citet{Hunter95} and \citet{McLaughlin05} found $\gamma \simeq 2$ in F555W. In none of the above investigations a proper core could be identified, such that the scale radius $a$ was in all cases found to be of the order of the resolution limit.

R136 does not appear to be kinematically relaxed \citep{Selman99}. In several studies evidence of dynamical substructure has been found, such as a ring of massive stars at about 2-3 pc radius \citep{Malumuth94, Brandl07}, or a shell of massive stars at a radius of 6 pc \citep{Hunter95} - structure which should be quickly erased by two-body relaxation. Moreover, the azimuthal density profile of R136 shows strong variations \citep{Campbell10}. Such substructure and asymmetry also causes bumps in the surface brightness profile at the corresponding radii \citep{Selman99, Malumuth94}, making the investigation of R136 difficult.

\section{Models}\label{Sec:Models}
\begin{figure*}
\centering
\framebox{
\includegraphics[width=70mm]{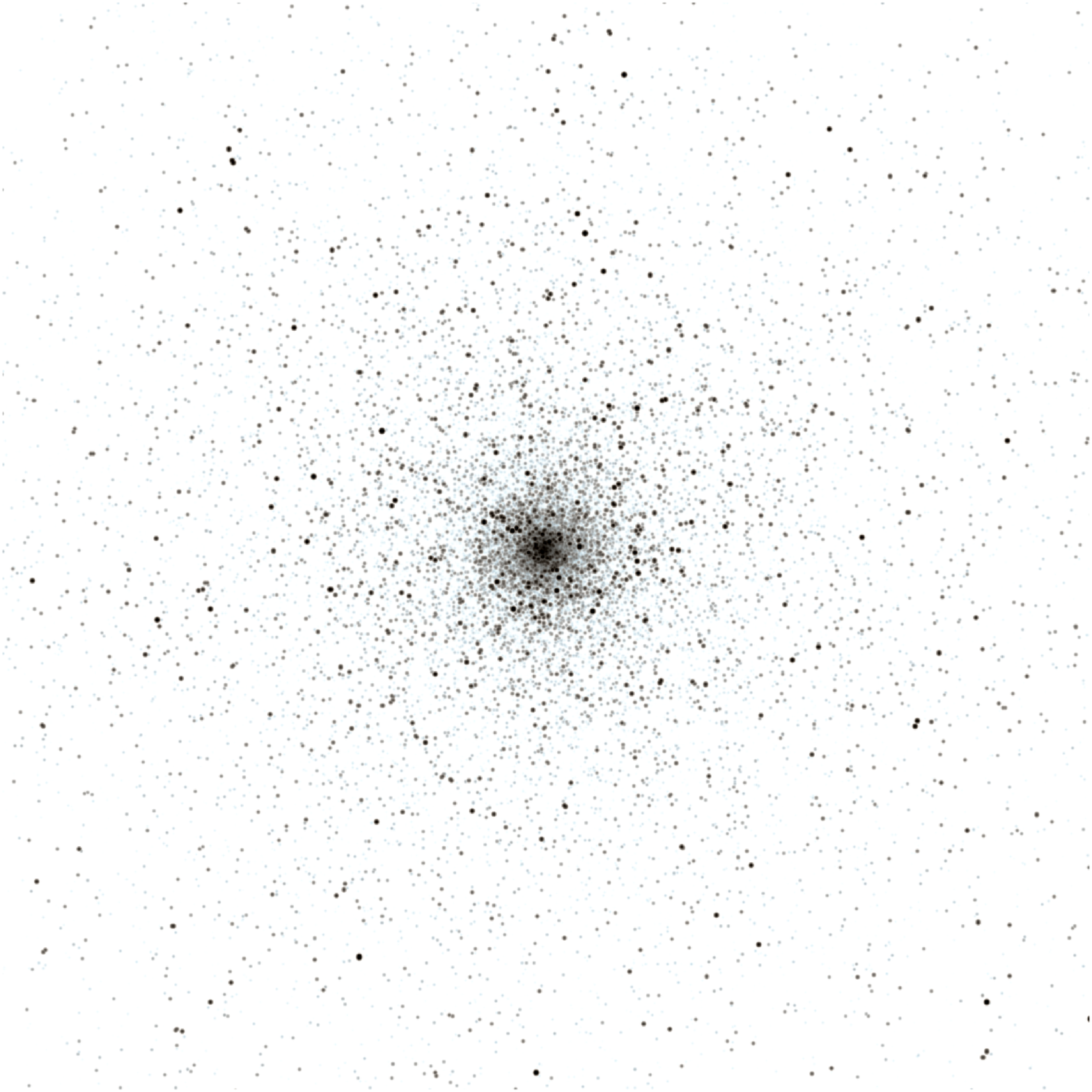}}
\framebox{
\includegraphics[width=70mm]{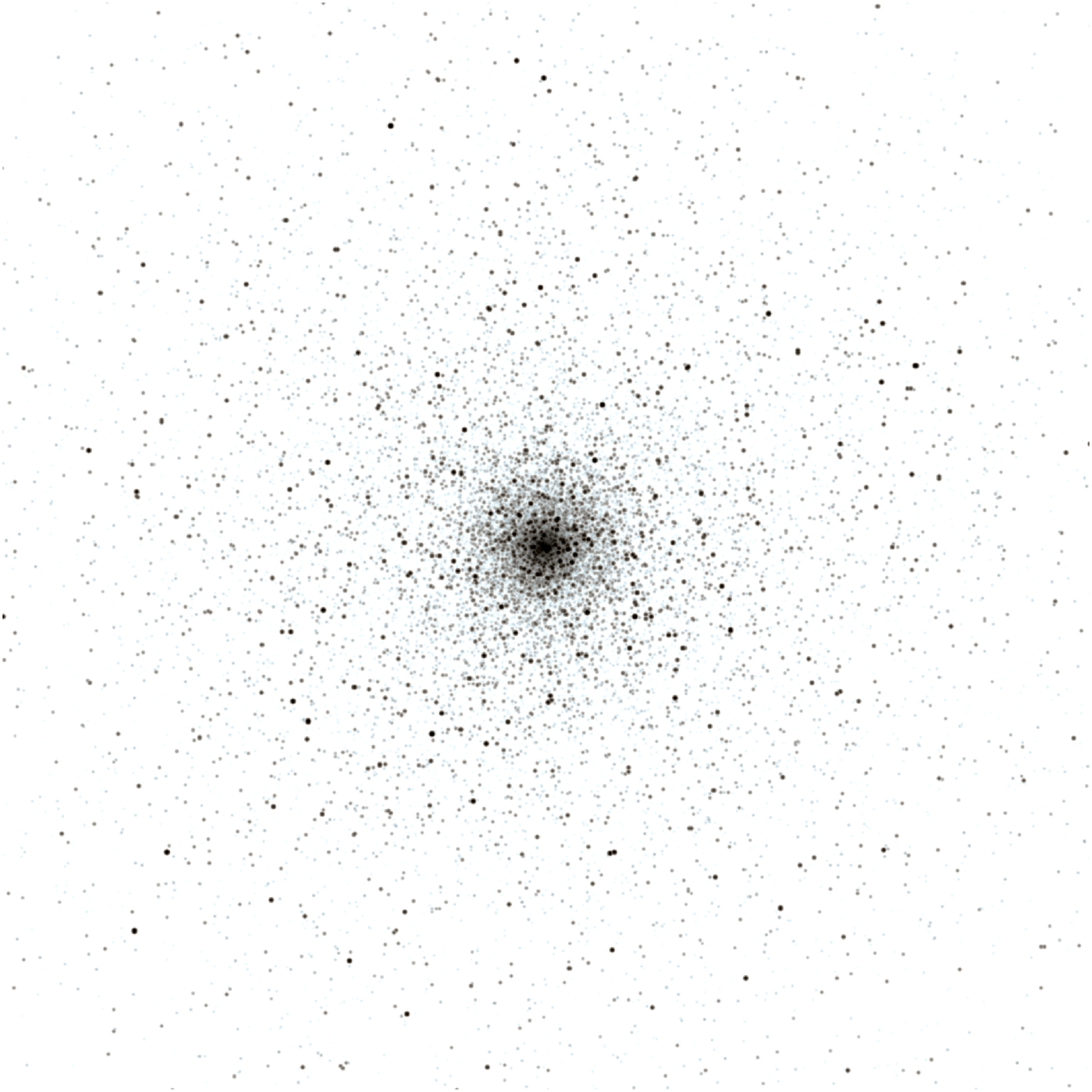}}\\
\framebox{
\includegraphics[width=70mm]{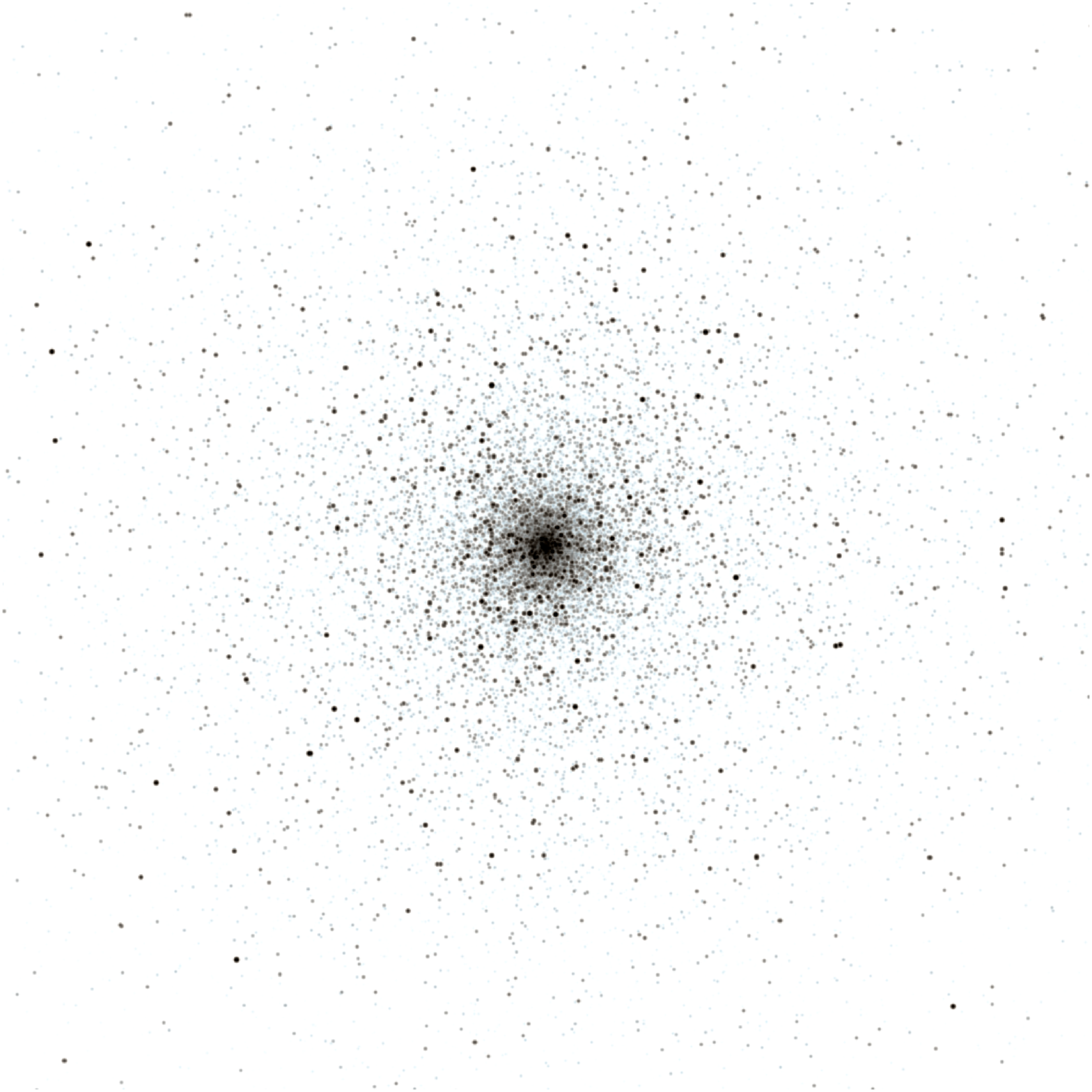}}
\framebox{
\includegraphics[width=70mm]{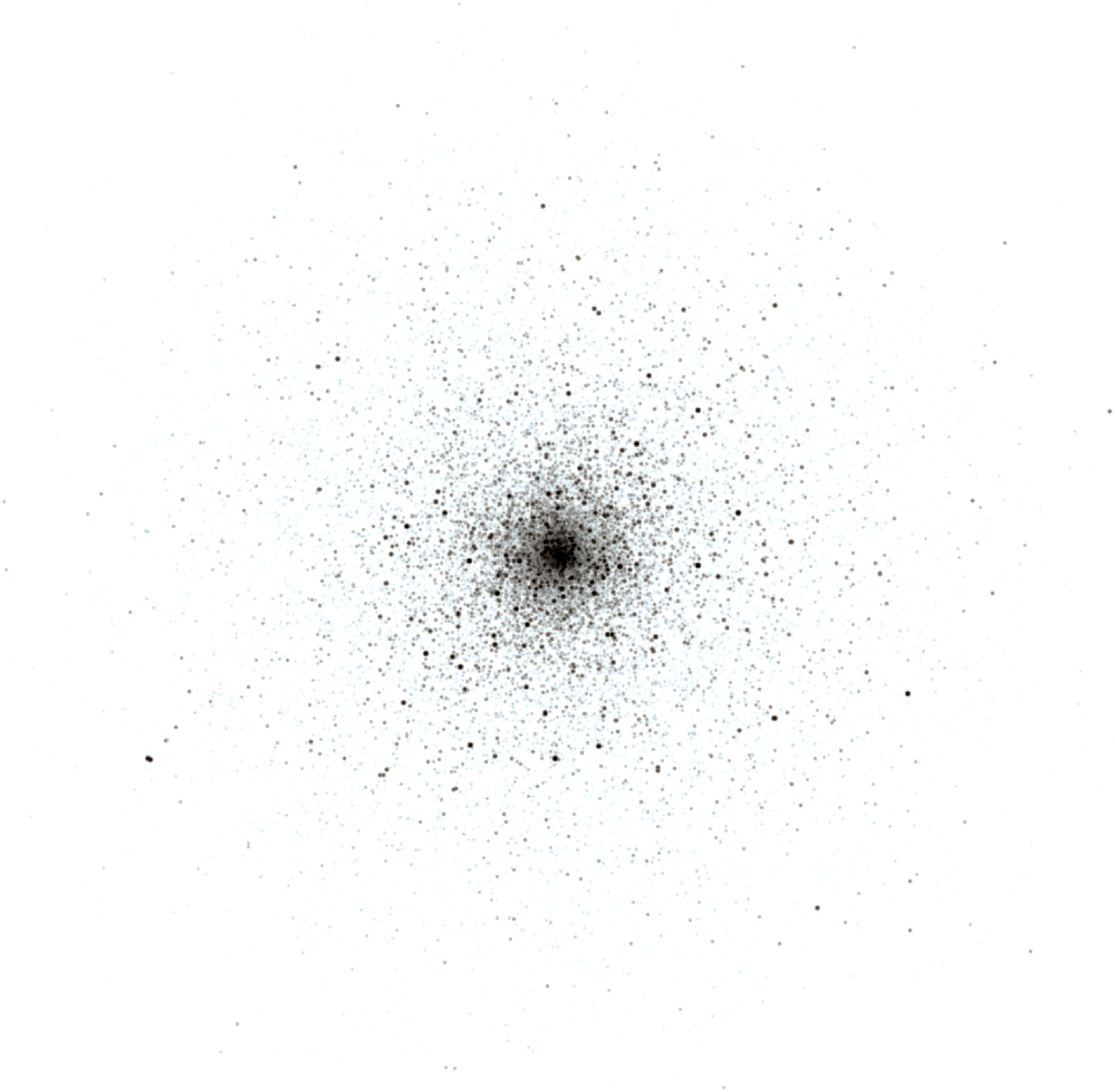}}\\
\framebox{
\includegraphics[width=70mm]{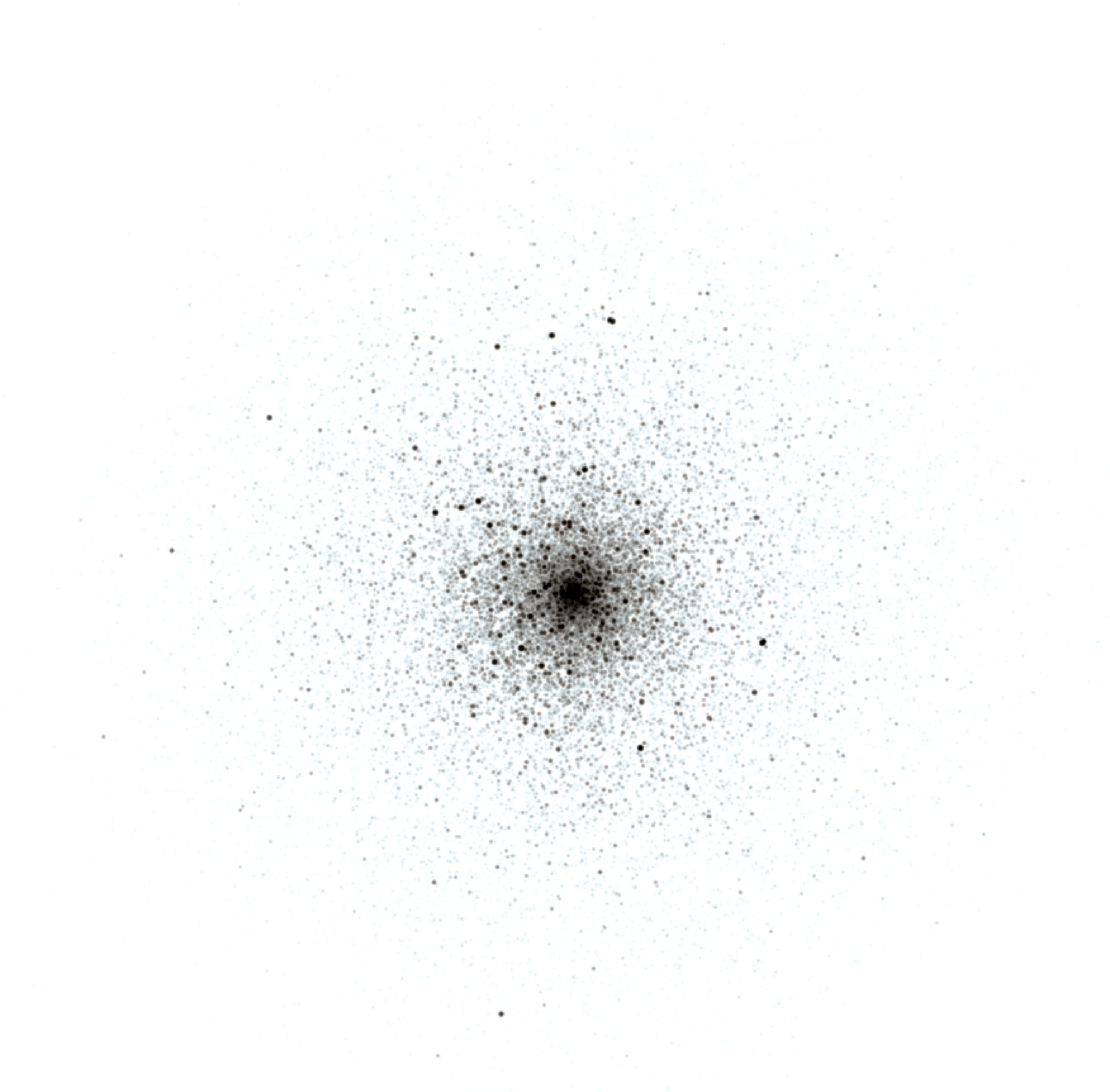}}
\framebox{
\includegraphics[width=70mm]{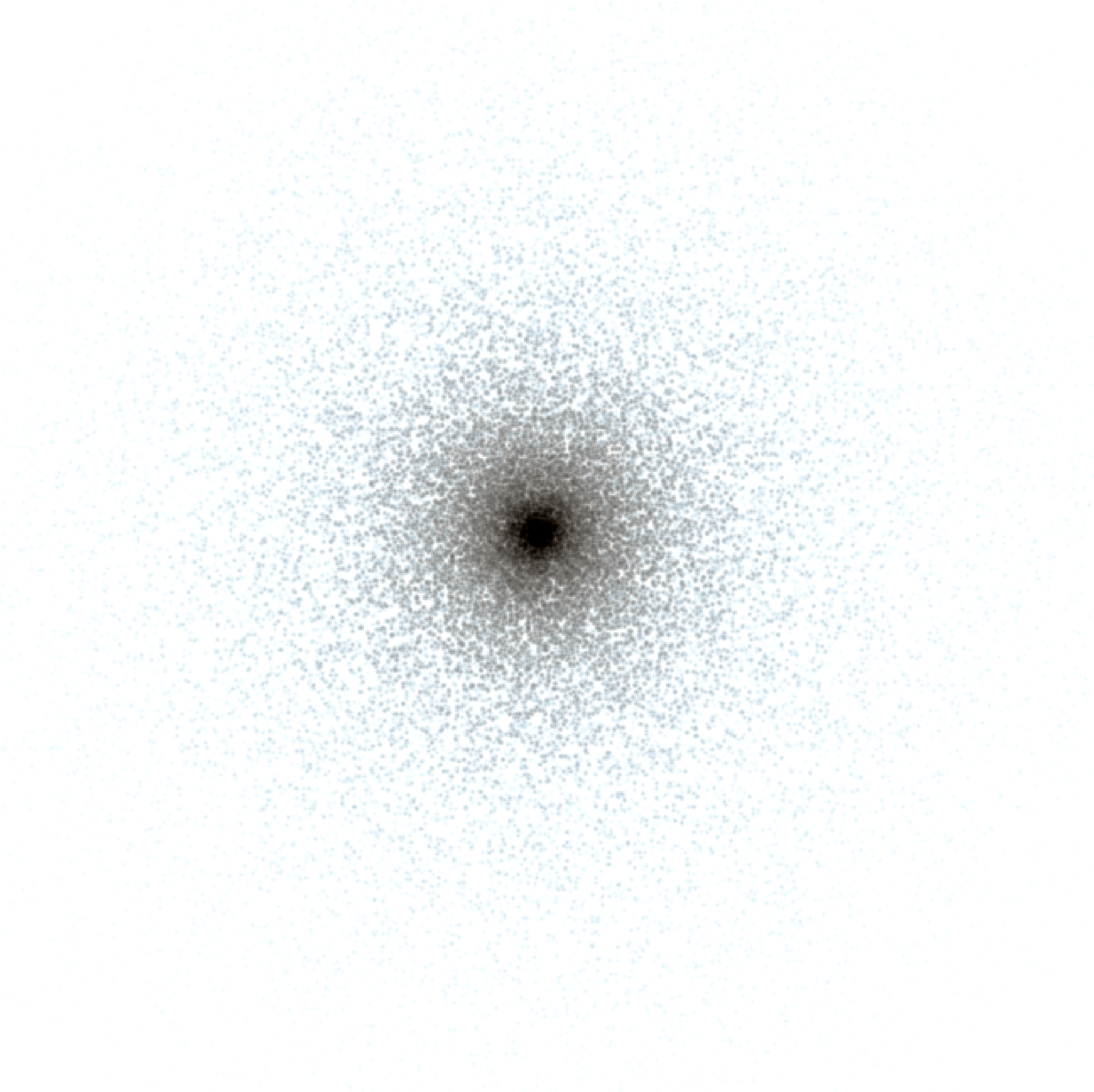}}\\
  \caption{Logarithmic intensity maps of 6 clusters models which where set-up with \textsc{McLuster}. The field-of-view is 20 pc $\times$ 20 pc. The clusters show different degrees of mass segregation. The mass segregation parameter, $S$, of the clusters is 0.0 (upper left), 0.4 (upper right), 0.7 (middle left), 0.9 (middle right), 0.95 (lower left), and 1.0 (lower right), respectively. All models follow the same mass profile and extend out to a radius of 20 pc.}
  \label{Segregation}
\end{figure*}
\begin{figure*}
\centering
\framebox{
\includegraphics[width=70mm]{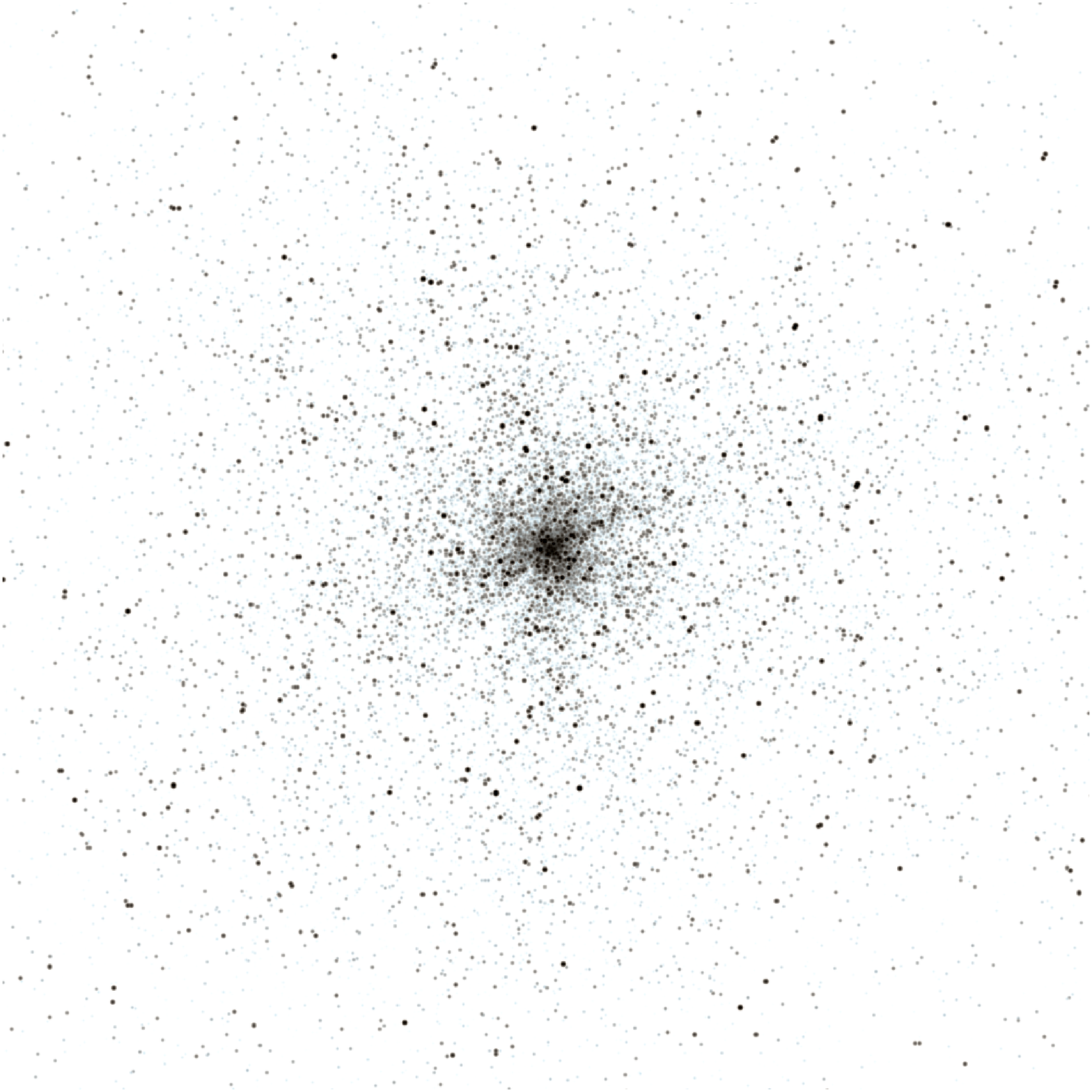}}
\framebox{
\includegraphics[width=70mm]{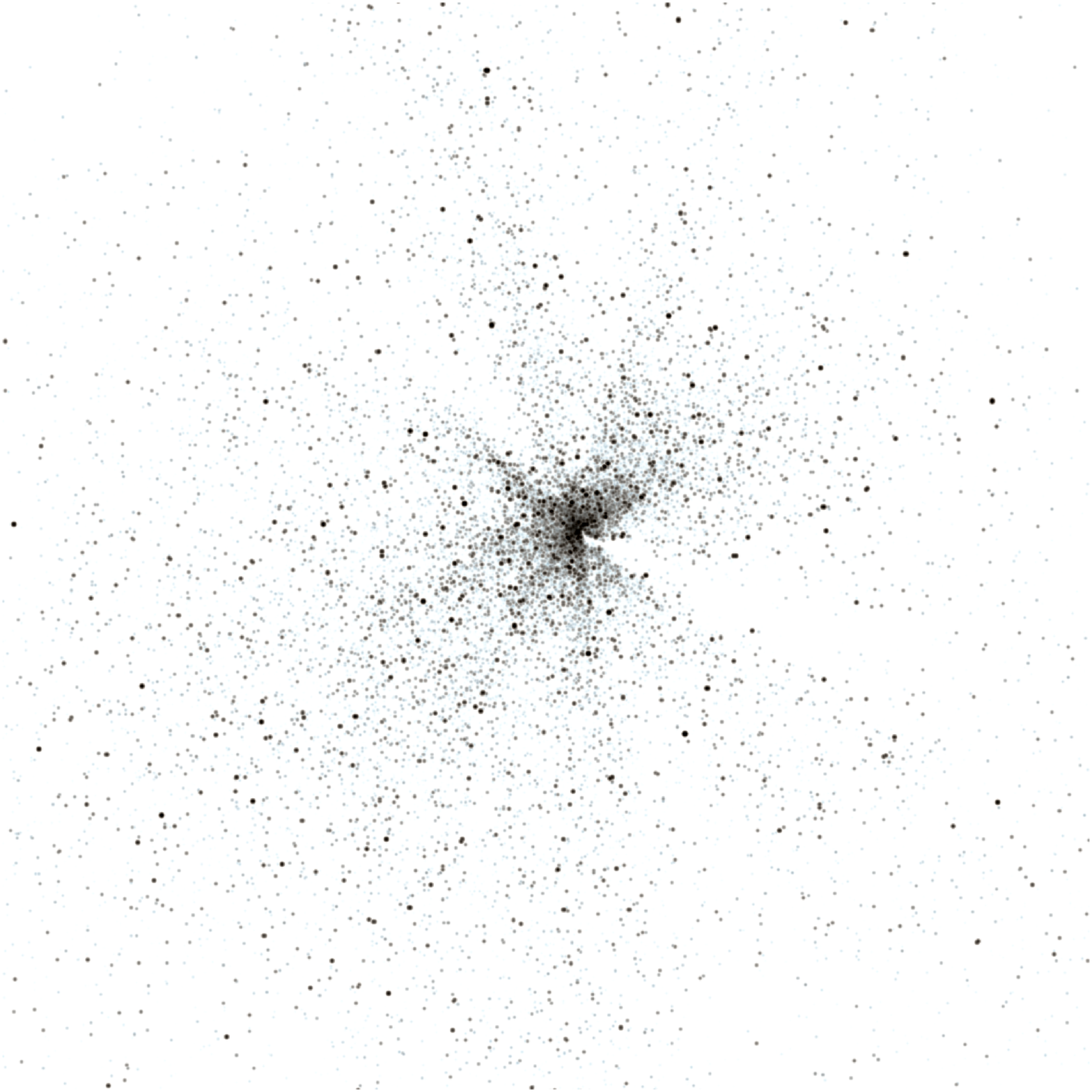}}\\
\framebox{
\includegraphics[width=70mm]{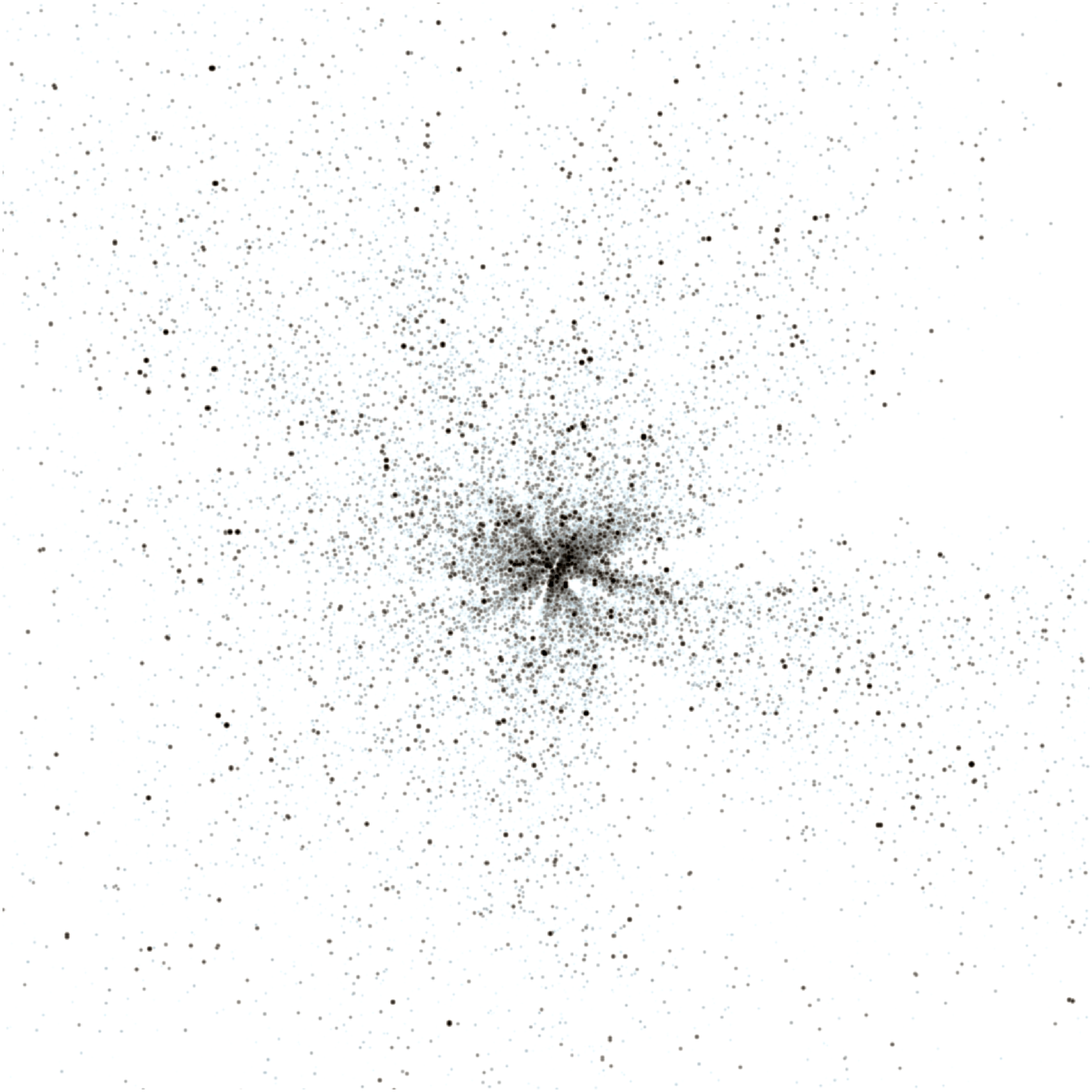}}
\framebox{
\includegraphics[width=70mm]{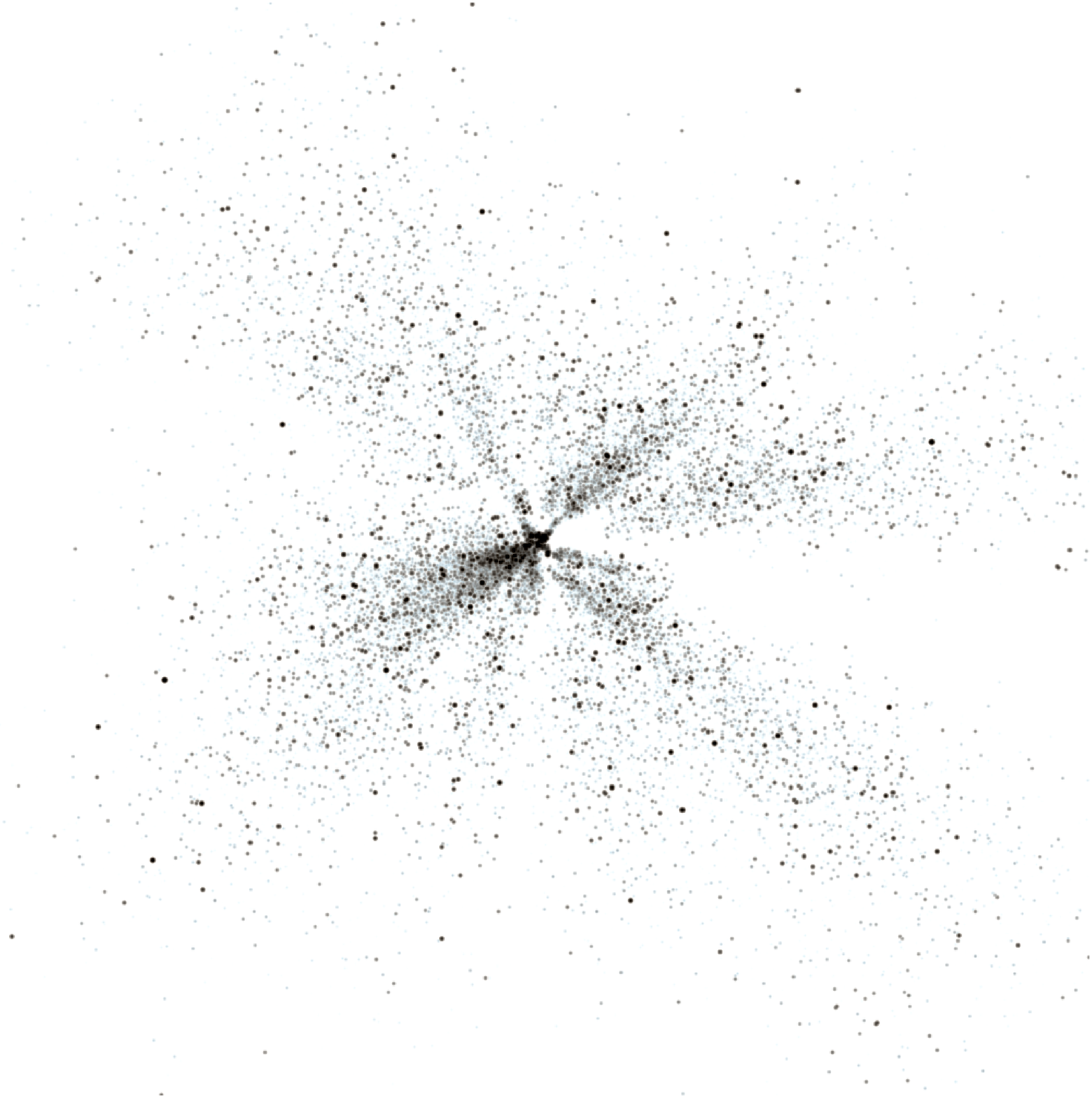}}
  \caption{Logarithmic intensity maps of 4 cluster models which where set-up with \textsc{McLuster}. The field-of-view is 20 pc $\times$ 20 pc. The clusters are not mass segregated ($S=0$) and show different degrees of substructure. The fractal dimension, $D$, of the clusters is 2.6 (upper left), 2.3 (upper right), 2.0 (lower left), and 1.6 (lower right), respectively. All models follow the same mass profile and extend out to a radius of 20 pc. Note that these clusters are not meant to look like real, existing objects but shall rather give a set of models with a smoothly increasing degree of substructuredness.}
  \label{Fractal}
\end{figure*}
To calibrate our methods for detecting mass segregation and substructure, we generated two sets of star cluster models with \textsc{McLuster}\footnote{\texttt{www.astro.uni-bonn.de/\~{}akuepper/mcluster/mcluster.html\\or  www.astro.uni-bonn.de/\~{}webaiub/german/downloads.php}\\A manual on our new publicly available cluster-initialisation code \textsc{McLuster} is provided in Appendix A.} with properties similar to those of R136 as seen today (see previous section). In addition, we produce one set of mass segregated models without binaries for comparison, and two more sets of models with substructure using different random seeds for estimating the stochastic scatter of the results.

All calibration models share the same basic properties, i.e. the clusters have a total mass of $10^5\msun$, consisting of about 170,000 stars drawn from the canonical \citet{Kroupa01a} IMF following a power-law index, $\alpha$, of 1.3 for stellar masses between $0.08\msun$ and $0.5\msun$, and $\alpha = 2.3$ for stellar masses larger than $0.5\msun$. As the upper stellar mass limit we chose $100\msun$ since stellar evolution models for higher masses are rather unreliable and therefore cannot be properly modelled by the stellar evolution routines we use. \citet{Crowther10} find 4 stars in R136 to exceed the mass of $165\msun$, though. They also estimate the total number of stars with initial masses above $100\msun$ to be about 14 within a radius of 5 pc. While this limitation may only have a negligible effect on most measures we test here, it may well have a significant effect on the colour-gradient method to detect mass segregation due to their high luminosities (see Sec.~\ref{Sec:Methods}).

The cluster stars have a metallicity of $Z=0.01$ and they were evolved from the zero-age main sequence to an age of 3 Myr using the \textsc{SSE} routine \citep{Hurley00} within \textsc{McLuster} (see Appendix A). Thus, the most massive stars have masses of about $80\msun$. This was done in order that the cluster stars have comparable colours like the stellar population of R136 which is especially important for detecting mass segregation with the colour-gradient method.

The binary fraction, $f_{bin}$, is 1.0 in all calibration models (except, of course, for the set of models without binaries), i.e. all stars are in binaries. The binaries were set up using ordered pairing for stars more massive than $5\msun$ following the method introduced by Oh \& Kroupa (in prep.), i.e. the most massive star is in a binary with the second-most massive star, the third with the forth, and so on. Stars with masses below this threshold were paired randomly. The value of $5\msun$ is somewhat arbitrary, but it rests on the observation that late-type stars with masses below $1-5\msun$ follow well defined, simple pairing rules (random pairing from the IMF, the initial period distribution function from \citet{Kroupa95b}, thermal eccentricity distribution), while massive stars with masses larger than about $10\msun$ tend to have similar component masses and shorter periods \citep{Kroupa11}.

The orbital elements of the binaries were generated using the \citet{Kroupa95b} period distribution and a thermal eccentricity distribution \citep{Kroupa08}. As shown in \citet{Kuepper08}, this results in a significant number of binaries which are too wide to be bound inside the very dense environment of our models. The mean kinetic energy, or 'dynamical temperature', of centre-of-mass particles in such a configuration is about 1 km$^2$s$^{-2}$. Assuming that all binaries with binding energies lower than this dynamical temperature are unbound or get disrupted quickly \citep{Heggie75}, the effective binary fraction would be about 0.5, but about 1.0 among the high mass stars which is consistent with recent observations \citep{Bosch09}, even though \citet{Crowther10} find the four most massive stars in R136 not to be in binary systems. But those few objects appear to be peculiar in many aspects which may well be due to the frequent strong gravitational encounters they must experience. Thus, they may be neglected here in this respect.

The density profile, $\rho (R)$, of the models was chosen to be an EFF profile (eq.~\ref{eq:EFF}, \citealt{Elson87}) with a scale radius, $a$, of 0.1 pc and a slope of $\gamma = 2.0$. The 2D density profile was deprojected within \textsc{McLuster} and used to generate the 3D cluster configuration (for details see Appendix A). The (infinitely extended) EFF profile was cut off at a radius of 20 pc, as we are mainly interested in the inner $\sim$10 pc. The central density, $\rho_0$, was fixed such that the total mass within this radius is $100,000\msun$.

To guarantee a good comparability between the different models, we chose the same random seed for all models such that the stellar populations in all of them are the same, and only the spatial distributions of the stars are different. Just the two additional sets of models with substructure have each a separate random seed to estimate the effect of stochasticity of the initialisation process on the results. 

The two calibration sets of models with binaries differ only in one parameter, one has a varying degree of mass segregation, and the other has a varying degree of fractal substructure. 
\begin{enumerate}
\item We produced 10 models with mass segregation values, $S$, ranging from 0.0 (unsegregated) to 0.9 in steps of 0.1, and another 10 models with values of $S$ from 0.91 to 1.0 (completely segregated) in steps of 0.01 (Fig.~\ref{Segregation}). For segregating the clusters we chose the method suggested by \citet{Baumgardt08a}, which preserves the desired (mass) density profile. We refer to Appendix A for details on how the intermediate steps between unsegregated and completely segregated clusters were set up with \textsc{McLuster}. In short: with a higher value of $S$, more massive stars get higher probabilities to be placed deep in the potential well of the cluster, i.e. in the centre. Lower values of $S$ correspond to more similar probabilities between high and low mass stars, i.e. more random distributions.
\item We generated $3\times15$ models with fractal initial conditions (Fig.~\ref{Fractal}). The fractal dimension, $D$, of the models was varied from 3.0 (non-fractal) to 1.6 (highly fractal) in steps of 0.1. Each set of 15 models has a different random number seed to measure the stochastic scatter between different realisations. For fractalizing the stellar distributions, we used the procedure described in Appendix A. In short: we set up a fractal distribution of stars within a unit-sphere following roughly the method of \citet{Goodwin04} and ``folded'' this distribution with the desired density profile. In this way, we end up with a radially concentrated but fractal distribution of stars. This is reminiscent of the filamentary and radially oriented structure of dense star-forming gas in contracting molecular cloud cores. Moreover, this way of producing a radial but fractal distribution is an important innovation for testing substructure-detection algorithms, since young clusters like R136 are neither purely radial nor purely fractal (see Sec.~\ref{Sec:R136}). 
\end{enumerate}

Additionally, we produce another set of mass segregated models like the one above but without binaries, to determine the influence of binary stars on the methods for detecting mass segregation.

\section{Methods}\label{Sec:Methods}
There have been several attempts to detect and quantify mass segregation and substructure in (young massive) star clusters. Here we are going to apply some of these techniques to our models of R136 to test their feasibility and to get some comparability among them. In a follow-up investigation we will apply some of these methods to $N$-body computations to follow the time evolution of mass segregation and substructure. Since some of the methods are computationally intensive when applied to a cluster with 170,000 stars, we apply a low mass cut-off at $1.1\msun$ for most methods which leaves us with about 15,000 stars. This furthermore reflects the common situation faced with observational data which often suffers from incompleteness and crowding. 

\subsection{Mass segregation}
We consider the following methods for the detection and quantification of mass segregation: 
\begin{enumerate}
\item In analogy to the work of \citet{Andersen09} on R136, we measure the mass function slope, $\alpha$, of stars above $1.1\msun$ at projected radii between 3 pc and 7 pc. In this radial range Andersen et al.~achieve reasonably high completeness levels. For comparison, we also do this for all stars in the cluster and for all stars within a projected radius of 3 pc from the cluster centre. The slope $\alpha$ we determine with the Modified Maximum Likelihood estimator (MML) of \citet{Maschberger08}. A standard deviation is estimated using 100 Monte Carlo subsets of stars. Since all models are set up with a mass function slope of 2.3 for stellar masses above $0.5\msun$, mass segregated clusters should show a steeper slope outside 3 pc and a shallower slope inside this radius. The choice of 3 pc is somewhat arbitrary but is meant to establish some comparability to the results achieved by \citet{Andersen09}.
\item We measure radial colour gradients following the methodology of \citet{Gaburov08}. From the stellar radii and corresponding effective temperatures, which are provided by the \textsc{SSE} routine \citep{Hurley00} in \textsc{Nbody6} as well as  \textsc{McLuster} (see Appendix), we compute $B$-, $V$- and $I$-band magnitudes for each star. We use the algorithm described in \citet{Flower96} to first compute the bolometric correction, $BC$, and the colour index, $B-V$. From this we derive the $V$- and $B$-band magnitudes, $M_V$ and $M_B$, respectively. Finally, we derive the $B-I$ colour using the relation of \citet{Natali94} and with this the $I$-band magnitude, $M_I$, as well as the $V-I$ colour. Mass segregated clusters should show a difference in $V-I$ colour in the inner part with respect to the outer part of the cluster. 
\item \citet{Allison09} suggest a method of detecting mass segregation using a minimum spanning tree (MST). Their measure, $\Lambda$, shows if a subset of stars is more concentrated compared to a random subset of the same size. It is computed with
\begin{equation}\label{eq:MST}
\Lambda = \frac{\ \overline{l}\ }{l}\pm \frac{\ \overline{\sigma}\ }{l},
\end{equation}
where $l$ is the length of the minimum spanning tree of the subset, $\overline{l}$ is the mean length of the MST of random subsets, and $\overline{\sigma}$ is the standard deviation of the distribution of MST lengths of random subsets. In a mass segregated cluster the most massive stars should show a $\Lambda$ well above 1 because they are more concentrated than the average subset of random stars. Here we are going to take only stars more massive than $1.1\msun$ into account, even though the computational expense of this method does barely depend on the total number of stars but on the size of the subset. This size, $N$, was practically limited to about 500 in our case, since the computation time scales with $\mathcal{O}(N^2)$ for Prim's algorithm which we used \citep{Prim57}. Moreover, we set the number of random subsets to 50, following the suggestion of \citet{Allison09}. 
\item \citet{Maschberger11} suggest a different method for quantifying mass segregation: by looking at the local stellar surface densities of stars. This measure has proven to be useful for detecting mass segregation in fractal structures which have not merged to a larger structure yet, but which may already be mass segregated \citep{Maschberger11}. It defines mass segregation differently than the MST measure: while the MST method measures how strongly the massive stars are grouped in relation to each other, the local surface density method measures how strongly the massive stars are grouped in relation to all stars. That is, the MST method measures how close the massive stars are to each other, whereas the local surface density method measures how close other stars are to the massive stars. We compute the projected stellar surface density around each star following \citet{Casertano85}, i.e. by measuring the radial distance, $R$, to its 6th nearest neighbour in projection and calculating the normalised local surface density as
\begin{equation}
\Sigma = \frac{6-1}{\pi R^2 \Sigma_{mean}},
\end{equation}
where $\Sigma_{mean}$ is the mean local surface density in the cluster. In a mass segregated cluster massive stars will have higher local densities than the average star. Thus, by normalising $\Sigma$ with the mean local surface density we get a dimensionless measure which should yield values larger than unity for mass segregated stars. For this method we also use only stars more massive than $1.1\msun$ since the computation of a neighbour list for $N$ stars scales with $\mathcal{O}(N^2)$ for a brute-force algorithm which we use here. To reduce the stochastic scatter, we bin the stars in mass bins of 500 stars starting from the most massive star. In this way, the $\Lambda$ and the local surface density measure are better comparable.
\end{enumerate}

\subsection{Substructure}
In order to detect and quantify substructure and asymmetry in our models we test three methods:
\begin{enumerate}
\item We look at the surface number density profile of massive stars. Like for the detection of mass segregation, we assume a reasonable completeness level above $1.1\msun$ and count only stars more massive than that. Inhomogeneities will appear as bumps and kinks in this kind of profile.
\item By measuring the azimuthal density profile (see e.g. \citealt{Gutermuth05}) we want to address possible asymmetries as was done by \citet{Campbell10} for R136. For this purpose we count the stars with masses above $1.1\msun$ within a projected radial distance of 7 pc in 20 azimuthal bins of 18 degree each. We quantify the asymmetry by computing the mean number density of the bins and the standard deviation. A comparable measure of asymmetry is then given by the normalised standard deviation, i.e. the standard deviation divided by the mean.
\item \citet{Cartwright04} suggest a measure for determining the degree of substructure, $Q$. For this method, the mean edge length, $\overline{m}$, of a minimum spanning tree connecting all stars in the cluster has to be measured and divided by the mean separation between the cluster stars, $\overline{s}$. For a homogeneous stellar distribution Cartwright \& Whitworth find a $Q$ value of about 0.8. A more substructured cluster should show a lower value of $Q$, whereas higher values of $Q$ indicate a radial distribution of the stars.
\end{enumerate}

\section{Results}\label{Sec:Results}
We test our methods with our calibration models in order to see how sensitive the various procedures are in determining mass segregation or substructure. In a follow-up investigation we aim at applying these methods to $N$-body models in order to see how mass segregation and substructure evolve with time in an R136-like configuration.

\subsection{Mass segregation}
\subsubsection{Mass function slope}
\begin{figure}
\includegraphics[width=84mm]{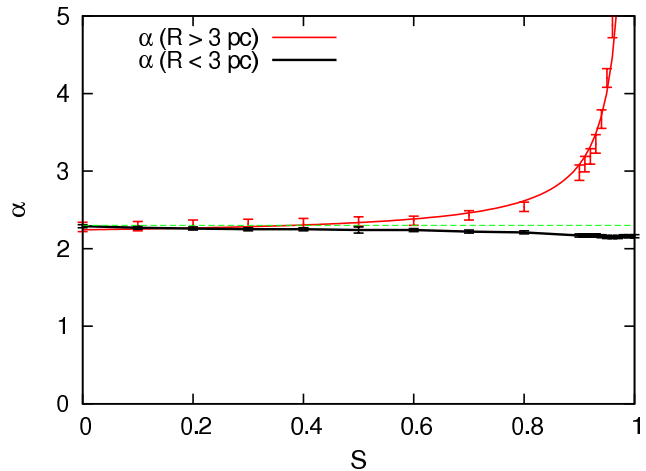}
  \caption{Mass function slopes of the stars with masses above $1.1\msun$ of the models with mass segregation and 100\% binaries among high mass stars. The points show the slopes for all stars within a projected radius of 3 pc (thick black) and for all stars outside this radius (thin red). The slopes were determined with the Modified Maximum Likelihood estimator of \citet{Maschberger08}. The uncertainties were estimated using a Monte Carlo approach. For comparison, the green dotted line gives the overall mass function slope of 2.3. The red solid line is Eq.~\ref{eq:alpha} with coefficients  $a = -0.09$ and $b = 2.15$. The difference in $\alpha$ becomes only evident for strongly mass segregated clusters with $S\ge0.7$.}
  \label{calibalpha}
\end{figure}
\begin{figure}
\includegraphics[width=84mm]{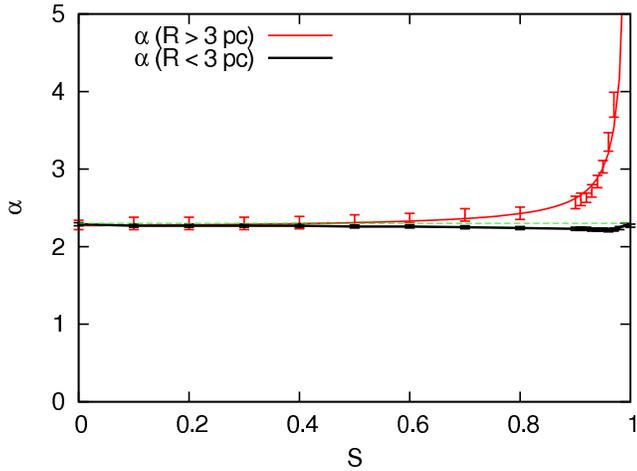}
  \caption{The same as Fig.~\ref{calibalpha} but for the models without binaries. The red solid line is Eq.~\ref{eq:alpha} with coefficients $a = -0.04$ and $b = 2.23$. The difference to the models with binaries comes from the way the models are set-up: binaries are treated as single particles with the sum of the component masses in the set-up process, i.e. the difference between the most massive and the least massive particle is larger in the case of a high binary fraction and such is the degree of mass segregation for any value of $S$.}
  \label{calibalphaF00}
\end{figure}
In Fig.~\ref{calibalpha} we show the results of the mass function slope determination for all stars above $1.1\msun$ within a projected radius of 3 pc and outside this radius. Also shown are the uncertainties of these values, which have been estimated using 100 Monte-Carlo subsets for each model. While there is barely any change in the measured mass function slope within 3 pc, at radii larger than 3 pc $\alpha$ changes significantly for high degrees of mass segregation, i.e. $S\ge0.7$. At radii larger than 3 pc the change in $\alpha$ follows a simple relation of the form 
\begin{equation}\label{eq:alpha}
\alpha(S) = a(S-1)^{-1}+b
\end{equation}
with fitted values $a = -0.09$ and $b=2.15$. 

We did the same for the models without binaries (Fig.~\ref{calibalphaF00}). The change in $\alpha$ with growing $S$ is less pronounced compared to the models with 100\% binaries among high mass stars ($a = -0.04$ and $b = 2.23$). This is due to the set-up process within \textsc{McLuster}, which first generates the binaries, replaces them by centre-of-mass particles with the combined mass of the two companion stars, and finally distributes those particles within the cluster before they are replaced by their constituent stars. With a high binary fraction the spread in masses between low-mass and high-mass particles is higher, and the number of particles is lower during the distribution process. Thus, the final degree of mass segregation is higher in the case of high binary fractions. For the binary free models, mass segregation becomes only significant for models with $S\ge0.8$.

\subsubsection{Colour gradient}
\begin{figure}
\includegraphics[width=84mm]{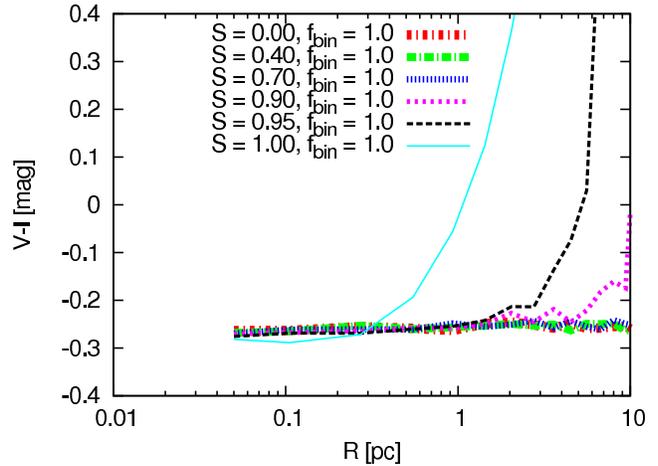}
  \caption{V-I colour profiles of the mass segregation models as suggested by \citet{Gaburov08} for the detection of mass segregation. Only very high degrees of mass segregation ($S\ge 0.9$) produce gradients larger than 0.1 mag within a radial range of about 10 pc. }
  \label{calibVI}
\end{figure}
\begin{figure}
\includegraphics[width=84mm]{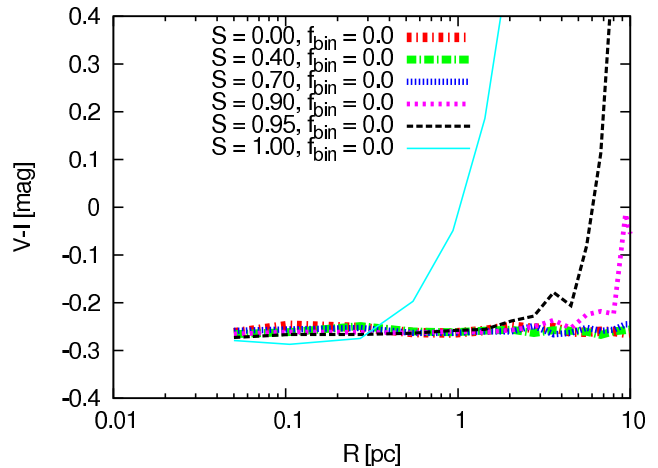}
  \caption{The same as Fig.~\ref{calibVI} but for the models without binaries. As for the mass function slope (Fig.~\ref{calibalphaF00}), the effect of growing $S$ is comparable to the case with the high binary fraction.}
  \label{calibVIF00}
\end{figure}
In Fig.~\ref{calibVI} we show the V-I colour profiles for the mass segregated models. A significant magnitude difference of more than 0.1 mag between the inner part of the clusters and the outer part is only observable for very high degrees of mass segregation ($S\ge0.9$). 

In Fig.~\ref{calibVIF00} the same is shown for the models without binaries. Like for the mass function slope, the effect is less pronounced due to the lower effective degree of mass segregation for models without binaries. In both cases, with and without binaries, mass segregation would only be detectable for clusters with mass function slopes $\alpha\ge3$ outside the core using this method on a R136-like young massive cluster.

It has to be kept in mind, though, that the adopted upper initial mass limit in our models is $100\msun$, whereas \citet{Crowther10} estimate the number of stars exceeding this limit to be of the order of 10. Such high-mass stars, which may even reach masses of up to $320\msun$ in R136 \citep{Crowther10}, will contribute significantly to the blue part of the spectrum, thus will make an observable colour gradient more likely if those stars are preferentially found near the cluster centre. 

\subsubsection{Minimum spanning tree}
\begin{figure}
\includegraphics[width=84mm]{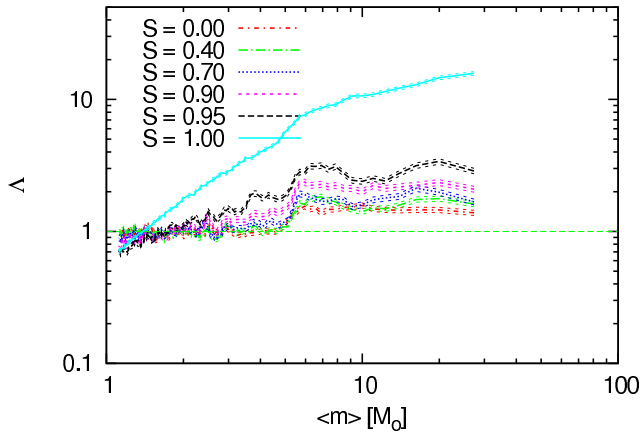}
  \caption{Minimum spanning tree (MST) measure, $\Lambda$, as suggested by \citet{Allison09}. The thicker lines show the normalised MST lengths for bins of 500 stars versus mean stellar mass, $<$$m$$>$. The thinner lines show the standard deviations from the mean of 50 random subsets of stars, which is a measure of the significance of the detections. All models show a jump at $5\msun$ which corresponds to the mass limit of binaries with similar mass companions. Below this threshold the binaries are paired randomly. The measure shows a slowly but continuously increasing degree of mass segregation for all models with $S\le 0.95$ and an extreme behaviour for the completely mass segregated model.}
  \label{calibMSTmean}
\end{figure}
\begin{figure}
\includegraphics[width=84mm]{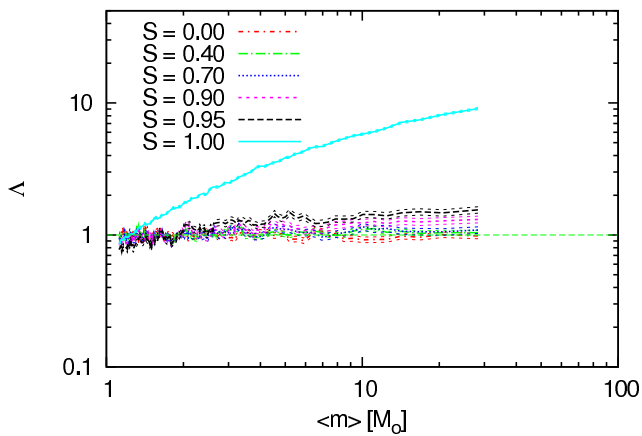}
  \caption{The same as Fig.~\ref{calibMSTmean} but for the models without binaries. The curves do not show the jump at $5\msun$ evident in the models with the high binary fraction. Only high values of $S\le0.9$ yield a significant signal. As in the case with binaries, the difference between complete mass segregation and lower values of $S$ is again high.}
  \label{calibMSTmeanF00}
\end{figure}
In Fig.~\ref{calibMSTmean} the minimum spanning tree measure, $\Lambda$, of \citet{Allison09} is shown for some of the mass segregated models. The mean MST length was determined for all stars in bins of 500 stars and divided by the mean MST length of 50 random subsets of 500 cluster stars each. The dotted lines give the standard deviation of the random subsets from this mean value (see eq.~\ref{eq:MST}). It demonstrates the large significance of the detected mass segregation in all clusters.

We see that all curves have a jump at $5\msun$, even the $S=0.0$ case, which is due to the binary component pairing routine in \textsc{McLuster}. We chose to pair massive O- and B-stars with similar mass companions, whereas all stars with masses less than $5\msun$ are paired randomly. This affects the MST length of the massive stars significantly. In Fig.~\ref{calibMSTmeanF00} we see that this jump disappears for the models without binaries. In both cases, most curves show a similar behaviour ($\Lambda$ mostly between 1 and 3) with a clear trend to higher $\Lambda$ for high values of $S$. Only the curve for $S=1.0$ significantly stands out from the rest. This is due to the fact that in the case of $S=1.0$ the high-mass stars have the smallest MST that can possibly be made. For values of $S$ smaller than 1.0 it becomes likely that larger edges are included in the high-mass MST such that its length grows rapidly.

\subsubsection{Local surface density}
\begin{figure}
\includegraphics[width=84mm]{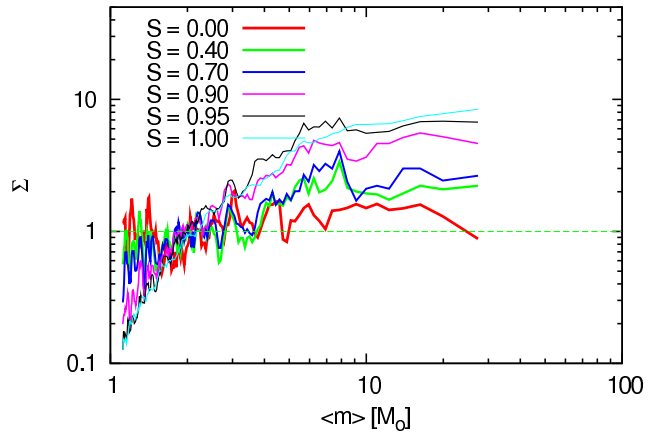}
  \caption{Local surface density measure, $\Sigma$ (eq.~3; \citealt{Maschberger11}). The lines show the normalised, median local surface density versus stellar mass. The lines were averaged with bins of 500 stars. This measure is only weakly affected by the binary pairing and shows a smooth transition from unsegregated to completely mass segregated models.}
  \label{calibNN500}
\end{figure}
\begin{figure}
\includegraphics[width=84mm]{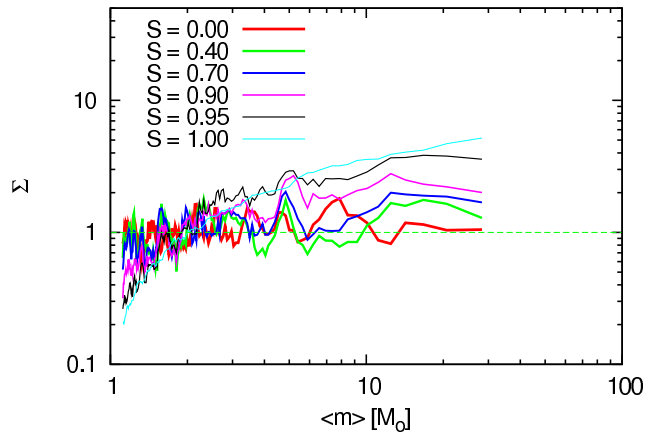}
  \caption{The same as Fig.~\ref{calibNN500} but for the models without binaries. On average the curves lie at lower values of $\Sigma$ and show a smaller scatter than for the models with binaries. }
  \label{calibNN500F00}
\end{figure}
In Fig.~\ref{calibNN500} we show the normalised local surface density, $\Sigma$, for bins of 500 stars. The curves look similar to the MST curves but appear to be less influenced by the ordered binary pairing.  Comparing Fig.~\ref{calibNN500} with Fig.~\ref{calibNN500F00}, which shows the same measure but for the clusters without binaries, shows that the binaries have indeed only a minor influence on the curves. In both cases, the $\Sigma$-values of the highest mass bins grow continuously with increasing $S$. Below $\sim10\msun$ the local surface density measure suffers from statistical variations, though.

In contrast to the MST measure, the local surface density measure  does not show a jump between the $S=0.95$ model and the completely mass segregated model. This is due to the fact that the local surface density of the high-mass stars is less influenced by outliers, since it is the average of 500 surface density values, whereas the high-mass MST length is a sum of 500 edge lengths where one outlier can make a large contribution.

\subsubsection{Comparison}
\begin{figure}
\includegraphics[width=84mm]{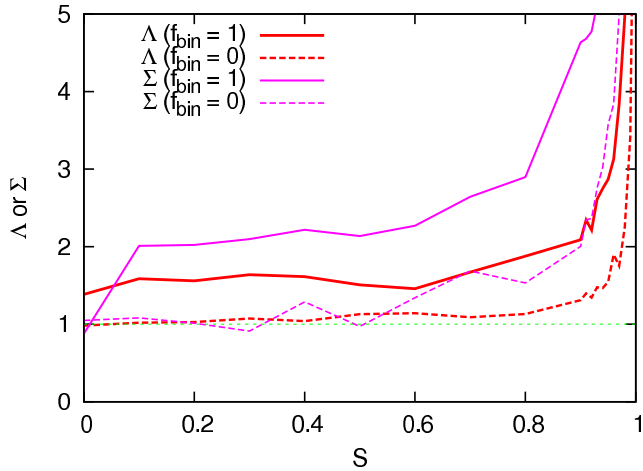}
  \caption{Summary of the methods for quantifying mass segregation for the uppermost mass bin (500 most massive stars) of all models with mass segregation (solid lines are models with binaries, dashed lines are for models without binaries). The $\Lambda$ is the MST measure and the $\Sigma$ is the local surface density measure. The ordinate gives the normalised MST length or the normalised local surface density, respectively. Note: whereas $\Lambda = \Sigma = 1$ signifies no mass segregation of the highest mass stars, $\Lambda =X$ is not equivalent to $\Sigma = X$ in general. }
  \label{calibMSTres}
\end{figure}
We have shown that the mass function slope method appears to be the easiest and most reliable way to detect mass segregation for a rich cluster like R136, even though it is not very sensitive to low degrees of mass segregation  and suffers from the arbitrariness of the choice of radius. That is, the results of this method depend on the radial range in which the mass function is measured and on the underlying mass function of the cluster. This complicates the interpretation of its results. 

The colour gradient method is also easy to work with but is very insensitive in an R136-like configuration. A further test including stars with very high initial mass ($\ge 100\msun$) would be very interesting, though, since, due to crowding and incompleteness effects, all but the colour gradient measure are almost impossible to compare with observational data of such a cluster.

The MST measure and the local surface density method do work for a R136-like cluster for non-seeing limited data, but are computationally much more demanding than the other two methods. In contrast to the others, both measures, MST and local surface density, allow not only to detect but also to quantify mass segregation. That is, their results automatically relate the behaviour of the massive stars to the other stars. In practice, the quantification is complicated by stochastic fluctuations.

The local surface density measure has the advantage that, once the neighbour density of each star is calculated, the stochastic fluctuations can be reduced easily by increasing the bin size which, in addition, increases its sensitivity. This is not possible with the MST measure, since new minimum spanning trees have to be constructed when the sample size is changed. With a sample size of 500 the local surface density measure shows a smoother behaviour in the very high-mass part, whereas the MST measure is, on average, smoother down to lower masses. This may complicate the interpretation of the local surface density measure when looking at the whole mass spectrum. Moreover, the MST measure has the advantage that the significance of its results is calculated simultaneously. Therefore, an estimate of the significance of the local surface density measure's results, like the standard deviation of the MST measure, should be constructed in order to make the local surface density measure more valuable.

In Fig.~\ref{calibMSTres} we compare the MST measure with the local surface density method. In this figure we only show the value of the uppermost mass bin, and show its dependence on the degree of mass segregation, $S$. Both measures show a steep rise for the highest values of $S$. Moreover, both measures are affected by a high binary fraction which may be due to the way we construct our binary population, since we pair massive stars above $5\msun$ preferentially with each other. Nevertheless, the effective degree of mass segregation should grow monotonically with increasing $S$. In this respect, the local surface density measure shows a more monotonic behaviour, even though it also suffers from statistical fluctuations (pink lines). It also gives a value of about 1 for the completely unsegregated cluster in the case of a high binary fraction (pink solid line). 

\subsection{Substructure}
\subsubsection{Radial density profile}
\begin{figure}
\includegraphics[width=84mm]{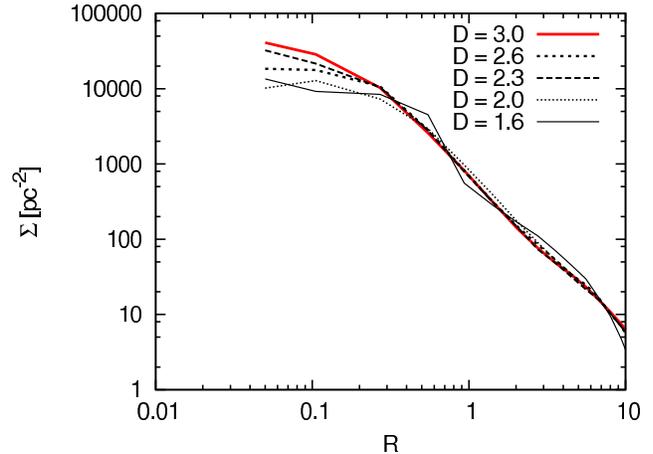}
  \caption{Radial number density profile for stars more massive than $1.1\msun$ of the clusters with fractal substructure. For lower values of $D$ the profiles deviate more strongly from the spherically symmetric case ($D = 3.0$). Only the model with the highest degree of substructure ($D=1.6$) shows a significant bump at about 0.5 pc and a kink in the power-law profile.}
  \label{calibSDP}
\end{figure}
In Fig.~\ref{calibSDP} we show the projected 2d radial number density profiles of the clusters with fractal substructure. For the profiles we only took stars more massive than $1.1\msun$ and put them into 20 logarithmically spaced bins between 0.05 pc and 20 pc radial distance. A cluster without substructure ($D=3.0$) is shown for comparison (red solid line). We see that for higher degrees of substructure the deviations from the spherically symmetric cluster grow. 

For the cluster with $D=1.6$, which has the highest degree of substructure in our sample, the radial profile shows a bump and a kink in the slope at about 0.5 pc. Depending on the radial range, a power-law fit of the profile would yield quite different results for such a cluster or the profile could be even interpreted as following a two-part power-law. For lower degrees of substructure, the radial density profile shows much less pronounced deviations making this plot rather unfeasible for quantifying the degree of substructure.

\subsubsection{Azimuthal density profile}
\begin{table*}
\centering
\caption{Azimuthal number density variations of the three sets of substructured models.  $\overline{\Sigma}$ gives the mean azimuthal number density of stars above $1.1\msun$ within a projected radial distance of 7 pc, whereas $\mbox{d}\Sigma$ gives the standard deviation from this mean.}
\begin{tabular}{cccccccccc}
\hline
 &\multicolumn{3}{c}{Set 1}       & \multicolumn{3}{c}{Set 2}      & \multicolumn{3}{c}{Set 3}\\
 $D$ & $\overline{\Sigma}$	[pc$^{-2}$] & $\mbox{d}\Sigma$	[pc$^{-2}$] & $\mbox{d}\Sigma/\overline{\Sigma}$ & $\overline{\Sigma}$	[pc$^{-2}$] & $\mbox{d}\Sigma$	[pc$^{-2}$] & $\mbox{d}\Sigma/\overline{\Sigma}$& $\overline{\Sigma}$	[pc$^{-2}$] & $\mbox{d}\Sigma$	[pc$^{-2}$] & $\mbox{d}\Sigma/\overline{\Sigma}$	\\
\hline										
 1.60 	& 55.2     & 33.4       & 0.61 & 46.5	& 34.1	& 0.74   &42.4		&32.9         &0.79\\
 1.70 	& 66.0     & 43.1       & 0.65 & 32.5	& 22.3	& 0.68	&49.1		&34.9		&0.71\\
 1.80	& 53.6     &  35.9      & 0.65 & 39.6	& 28.4	& 0.73	&52.5		&37.7		&0.72\\
 1.90 	& 54.4     &  29.9      & 0.55  & 40.6	& 20.2	& 0.50	&47.2		&25.3	        &0.54\\
 2.00 	& 41.7     &  24.3      & 0.58  & 38.1	& 17.6	& 0.47	&42.7		&25.7		&0.61\\
 2.10 	& 54.1     &  22.2      & 0.41  & 49.3	& 31.6	& 0.64	&36.7		&17.4		&0.48\\
 2.20 	& 46.1     &  18.5      & 0.40  & 45.3	& 22.2	& 0.49	&52.8		&20.6		&0.39\\
 2.30 	& 46.7     &  18.3      & 0.39  & 51.0	& 17.5	& 0.36	&57.4		&22.3		&0.39\\
 2.40 	& 45.5     &  18.6      &  0.41 & 47.8	& 19.2	& 0.40	&47.7		&12.3		&0.25\\
 2.50 	& 52.4     &  14.3      &  0.27 & 44.4	& 14.4	& 0.33 	&55.4		&15.6		&0.29\\ 
 2.60 	& 49.7     &  11.6      &  0.23 & 48.4	& 10.1	& 0.21	&40.7		&11.4	        &0.28\\
 2.70 	& 43.9     &  10.3      &   0.23 & 46.9& 8.70	& 0.19	&43.8		&11.3		&0.26\\
 2.80 	& 44.2     &   8.02     &   0.18 & 46.4& 14.7	& 0.32	&44.4		&8.19		&0.19\\
 2.90 	& 46.0     &   6.16    &    0.13 & 47.5& 7.25	& 0.15	&49.8		&6.51         &0.13\\
 3.00 	& 47.7    &   3.20    &    0.07 & 47.0	& 3.30	& 0.07	&47.1		&3.98		&0.08\\
\end{tabular}
\label{table2}
\end{table*}
\begin{figure}
\includegraphics[width=84mm]{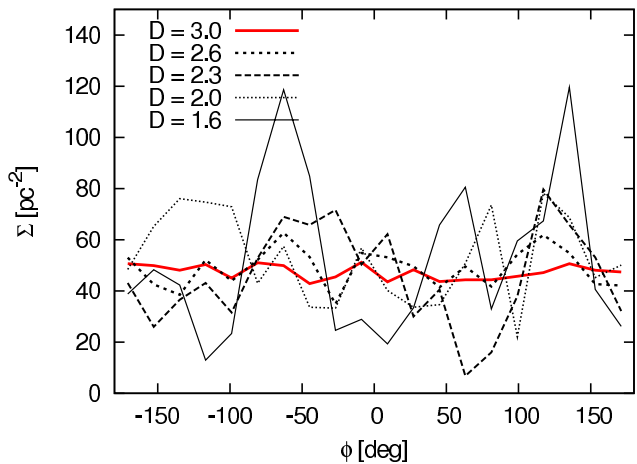}
  \caption{Projected azimuthal number density profile for stars more massive than $1.1\msun$ of the clusters with fractal substructure. The density was measured in 20 bins of 18 deg for all stars within a projected radial distance of less than 7 pc from the cluster centre. The azimuthal variations grow for lower values of $D$, i.e. for higher degrees of substructure.}
  \label{calibADP}
\end{figure}
\begin{figure}
\centering
\includegraphics[width=35mm]{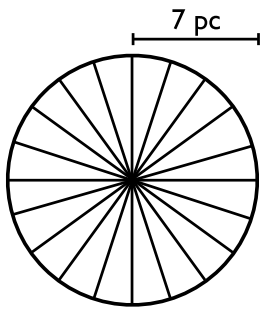}
  \caption{Sketch of how the projected azimuthal number density profile is determined. The density is measured in 20 bins of 18 deg for all stars within a projected radius of 7 pc from the cluster centre.}
  \label{sketch}
\end{figure}
\begin{figure}
\includegraphics[width=84mm]{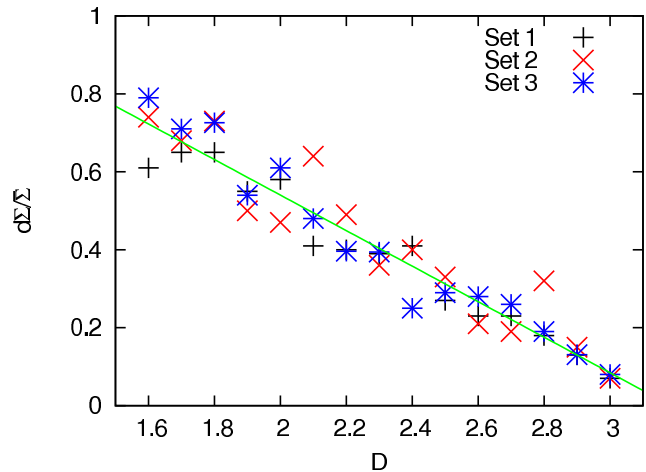}
  \caption{Relative azimuthal variations of the clusters with fractal substructure. Shown are three different sets of models each with a different random seed.} The solid line gives a linear fit to all the data (Eq.~\ref{eq:sigma}) with a slope of -0.46 and an intercept of 1.45.
  \label{calibdADP}
\end{figure}
In Fig.~\ref{calibADP} we show the azimuthal density profiles of the same clusters as in Fig.~\ref{calibSDP}. For this purpose, we used again only stars with masses above $1.1\msun$. Moreover, we counted only stars within a projected radius of 7 pc and put them into 20 bins of 18 degree width each (see Fig.~\ref{sketch}). The bins were then divided by the covered area in pc$^2$. Here, the different degrees of substructure are more apparent. The spherically symmetric model (red solid line) shows only some minor statistical fluctuations, whereas the model with the highest degree of substructure shows a difference of about a factor 6 between the bin with the fewest and the bin with the most stars. The mean azimuthal density is for all clusters about the same (see also Tab.~\ref{table2}).

In Fig.~\ref{calibdADP} the relative azimuthal variations for all three sets of substructured models are shown, that is, the standard deviation of azimuthal densities from the mean azimuthal density, $\mbox{d}\Sigma$, divided by the mean, $\overline{\Sigma}$. From the different sets of models which are generated with three different random seeds we can see that the generation of models with substructure is a quite stochastic process. But, as expected, the relative azimuthal variation grows with decreasing values of $D$. The growth follows a simple relation of the form 
\begin{equation}\label{eq:sigma}
\frac{\mbox{d}\Sigma}{\overline{\Sigma}}(D) = a\,D+b,
\end{equation} 
with $a\simeq -0.46$ and $b\simeq 1.45$. 

The azimuthal density profile appears to be a good measure for substructure in star clusters which allows a reasonable quantification.

\subsubsection{$Q$ parameter}
\begin{figure}
\includegraphics[width=84mm]{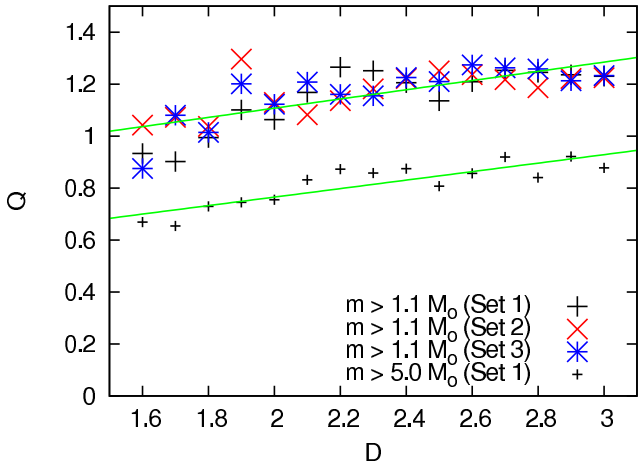}
  \caption{Cartwright \& Whitworth's $Q$ parameter of the clusters with fractal substructure. Computation of this parameter is computationally too demanding when all stars are taken into account and Prim's algorithm is used for the computation of the MST length (see text). But when calculating it for only a subset of the most massive stars, the absolute value of the $Q$ parameter seems to depend on the number of stars in the subset, which is most probably due to the binary population. The lower solid line gives a linear fit to the $Q$ parameters for the subsets containing all stars above $5\msun$ (Eq.~\ref{eq:Q}) with a slope of 0.16 and an intercept of 0.44. In contrast, the upper solid line is a fit to the subsets containing all stars with masses larger than $1.1\msun$ (about $15000$ stars) from all three sets of models. The slope here is 0.18 and the intercept 0.75.}
  \label{calibQ}
\end{figure}
In Fig.~\ref{calibQ} we show the $Q$ parameter as defined by \citet{Cartwright04} for our models with fractal substructure. The computation of this parameter necessitates construction of a minimum spanning tree of all cluster stars. As stated above, this is computationally very expensive for a cluster of 170,000 stars when using Prim's algorithm. Even for a subset of about 15,000 stars with masses larger than $1.1\msun$ the computation of this parameter takes several hours on a regular workstation. One would have to switch to a more sophisticated algorithm to use the full set of stars which is beyond the scope of this paper. We therefore compute the $Q$ parameter for different sizes of subsets with different low-mass cut-offs.

We compute $Q$ by measuring the mean separation between the cluster stars, $\overline{s}$, and the mean edge length, $\overline{m}$, of a minimum spanning tree connecting all stars in the subset, such that
\begin{equation}
Q = \frac{\ \overline{m}\ }{\overline{s}}. 
\end{equation}
Cartwright \& Whitworth find values between 0.45 for highly fractal configurations (fractal dimension $D=1.5$) and 0.8 for homogeneous configurations ($D=3.0$). Values above 0.8 and up to about 1.5 they find for models with a radial density gradient ($\rho_{3d} \propto r^{-3}$). But, in contrast to their investigation, the models in our sample have a radial density gradient  ($\rho_{2d} \propto R^{-2}$) and at the same time fractal substructure ($1.6 \le D \le 3.0$). Therefore, we expect the value of $Q$ of our models to be of order of the values Cartwright \& Whitworth find for the radial clusters, but also to show some variability around this value depending on the degree of substructure. Moreover, our models contain binaries which are paired non-randomly, i.e. the massive stars above $5\msun$ are paired with similar mass companions (see Sec.~\ref{Sec:Models}) which further complicates the interpretation of $Q$.

Interestingly, as can be seen in Fig.~\ref{calibQ}, the $Q$ values of our models depend on the size of the subset. For a small subset containing only the most massive stars with masses larger than $5\msun$ (about 2500 stars) the values almost agree with the findings of Cartwright \& Whitworth for purely fractal clusters but are shifted about 0.1 upwards. The more stars are taken into account the more the absolute values of the $Q$ parameter are shifted to larger values. This is due to the binaries in our clusters: when taking all stars above $5\msun$ we only have massive binaries in our sample, as the massive stars are paired with each other. When going to a lower mass limit we add stars to the sample that need not be in binaries, i.e. whose binary companions need not to be within the sample. This increases the mean edge length $\overline{m}$, whereas it barely affects the mean separation $\overline{s}$, such that $Q$ becomes larger. 

We fit a linear function to the values of the form
\begin{equation}\label{eq:Q}
Q(D) = a\,D+b,
\end{equation} 
where $a=0.16$ and $b=0.44$ for the small subsets with $m>5\msun$, whereas $a=0.17$ and $b=0.75$ for the large subsets with $m>1.1\msun$ (about 15000 stars) from all three sets of models. Subsets with sizes in between these two yield intermediate values (e.g. $a=0.21$ and $b=0.62$ for a subset with $m>1.9\msun$, i.e.  about 7500 stars).

To which exact values $Q$ will converge when more stars are taken into account cannot be checked easily but the growth in $Q$ with increasing sample size falls off strongly such that the overall values should be close to the values of our largest samples. Anyway, the question remains how useful such a quantification of substructure is. To follow the evolution of substructure with time in a single $N$-body computation seems possible with this method. But it is time consuming, unless a highly sophisticated algorithm is used.

\section{Summary and conclusions}\label{Sec:Conclusions}
Here we introduce for the first time the code \textsc{McLuster}, which is a publicly-available tool for initialising star cluster models and binary-rich stellar populations. 

Moreover, we have tested and calibrated several methods for detecting and quantifying mass segregation and substructure in non-seeing limited star cluster data. We applied these methods to models of the young massive cluster R136 which is the only starburst site in the Local Group. Accessing this cluster and measuring its degree of mass segregation and substructure at its current age of about 3 Myr is of importance because it is the only object which can be resolved into single stars and at the same time can give insight to star formation conditions in starburst sites. Moreover, due to its size and mass we can assume that the present-day conditions in R136 can still yield insights to the conditions 3 Myr ago, as the dynamical time of this configuration is comparably long (about $0.3$ Myr), i.e. the cluster is dynamically young \citep{Portegies10}. Finally, its moderate mass of about $10^5\msun$ makes it accessible by means of $N$-body investigations. But for understanding the development of these two quantities, mass segregation and substructure, easily quantifiable and little time consuming methods have to be found.

We compare 4 different methods for quantifying mass segregation and 3 methods for quantifying substructure from the literature (Sec.~\ref{Sec:Methods}). The former we quantify by comparing the mass function slope of massive stars, the radial colour profile \citep{Gaburov08}, the minimum spanning tree measure \citep{Allison09, Allison10} and the local surface density measure \citep{Maschberger11}. We quantify substructure by looking at the projected radial number density profile, the projected azimuthal density profile \citep{Gutermuth05, Gutermuth08} and by calculating the $Q$ parameter \citep{Cartwright04}. For this purpose we set up star cluster models with different degrees of mass segregation and substructure using the new publicly available code \textsc{McLuster} (Sec.~\ref{Sec:Models}). 

We find that the methods for detecting mass segregation often yield ambiguous results (Sec.~\ref{Sec:Results}). From the four methods we compare, the mass function slope seems to be the simplest and most reliable measure for detecting mass segregation, even though it suffers from a somewhat arbitrary choice of radius, and significant detections of mass segregation are only possible for high degrees of mass segregation. The radial colour profile only yields significant results for the highest (and rather unrealistic) degrees of mass segregation for a configuration like R136. Both measures cannot handle substructure, i.e. clusters which are not radially symmetric. In contrast to that, the minimum spanning tree measure and the local surface density measure can handle substructured clusters \citep{Maschberger11}, but their results are often ambiguous, and computationally they are much more demanding. Moreover, the minimum spanning tree method is strongly influenced by a high binary fraction. On the other hand, it has the unique advantage that the significance of its results is readily given. The local surface density measure is less influenced by binaries, and with careful calibration it can be a very sensitive method for detecting mass segregation. We recommend, though, that a measure of its significance similar to the standard deviation of the MST measure should be constructed.

For quantifying substructure we are left with the projected azimuthal density profile since the projected radial density profile only shows significant deviations from the spherical symmetric case for extremely substructured configurations. Such a cluster with a high degree of substructure can show strong bumps and kinks in its projected radial profile but those are difficult to quantify. The projected azimuthal density profile is a reliable measure of substructure (Fig.~\ref{calibdADP}). We suggest to compute the mean projected azimuthal density and the normalised standard deviation from this mean to get a useful measure. The $Q$ parameter is also sensitive to substructure but is computationally very demanding, when using a standard algorithm like Prim's algorithm, as a minimum spanning tree for all cluster stars has to be constructed. If only a subset of stars is taken account for its computation, its absolute value shows a dependence on the number of stars in this subset which is due to the binary population we adopted. Thus, the $Q$ parameter is only of limited use in such configurations, unless a much more sophisticated algorithm like Kruskal's algorithm is used (see e.g. \citealt{Lomax11}).

Finally, we have to conclude that most of the methods presented in this work will likely yield ambiguous results when applied to observations of young massive clusters, due to crowding and incompleteness effects, as has similarly been found by \citet{Ascenso09}. Also the distance of such clusters and heavy, variable extinction will add further difficulties. But even for non-seeing limited data of (young) massive clusters, like $N$-body data, some of the methods will face great computational difficulties due to the large number of stars involved.

\section*{Acknowledgments}
The authors are grateful to Sverre Aarseth for making his \textsc{Nbody} codes accessible to the public and for many useful discussions. Moreover, they would like to thank Douglas Heggie, Ladislav {\v S}ubr and Jarrod Hurley for allowing the usage of their codes within \textsc{McLuster}. The \textsc{McLuster} code makes use of Numerical Recipes routines \citep{Press86}. AHWK kindly acknowledges the support of an ESO Studentship and through the German Research Foundation (DFG) project KR 1635/28-1. H.B. acknowledges support from the Australian Research Council through the Future Fellowship grant FT0991052.

\input{appendix.tex}

\bsp

\label{lastpage}
\end{document}

%% file: appendix.tex
\appendix
\begin{table*}
\caption{Overview of available command line options in \textsc{McLuster} with brief descriptions. For details on the available choices see the corresponding paragraphs. Also given are the possible ranges and the default values (which can be permanently changed within \texttt{main.c}).}
\begin{center}
\begin{tabular}{cccl}
Option & Range &  Default & Meaning\\ \hline
      -N & $0<N<Nmax$ & 0 & Number of cluster stars (specify either $N$ or $M$)\\                                 
      -M & $M>0$ & 1000 &Mass of the cluster ($\msun$; specify either $N$ or $M$)\\           
      -P & 0/1/2/3/-1& 0 &Density profile (Plummer/King/\v{S}ubr/EFF/homogeneous sphere)\\             
      -W & 1.0--12.0 & --- & $W_0$ parameter for the King model (only $P=1$, ignored if $P\neq 1$)\\                       
      -R & $R>0$ & 0.8 & Half-mass radius in parsec (ignored for $P = 3$)\\         
      -r & $0<r<c$ & --- & Scale radius of the EFF template in parsec (only $P=3$, ignored if $P\neq 3$)\\         
      -c & $c>r$ & --- & Cut-off radius of the EFF template in parsec (only $P=3$, ignored if $P\neq 3$)\\       
      -g & $g>1.5$ & --- & Power-law slope of the EFF template (only $P=3$, ignored if $P\neq 3$)\\           
      -S & 0.0--1.0 & 0.0 & Degree of mass segregation (0.0 means no segregation; $S<1.0$ for $P=2$)\\
      -D & 1.6--3.0 & 3.0 & Fractal dimension (3.0 means no fractality)\\          
      -T & $T>0$ & 100.0 & \textsc{Nbody4/6} computation time in N-body units\\                            
      -Q & $Q\ge0$ & 0.5 & Virial ratio ($Q = 0.5$ for virial equilibrium)\\                                     
      -C & 0/1/3 & 3 & Output type (\textsc{Nbody6}/\textsc{Nbody4}/ASCII table)\\    
      -A & $A>0$ & 2.0 & \textsc{Nbody4/6} adjustment time in N-body units (e.g. \citealt{Heggie03})\\                            
      -O & $O>0$ & 2.0 & \textsc{Nbody4/6} output time in N-body units\\                           
      -G & 0/1 & 0 & Use GPU with \textsc{Nbody6} (no/yes)\\                   
      -o & --- & test & Output name of the cluster model\\                     
      -f & 0/1/2 & 1 & IMF (no mass function/Kroupa IMF/user defined)\\
      -a & $a<0$ & --- & IMF slope for a user defined IMF, may be used multiple times, from low to high mass\\       
     -m & $m>0$ & --- & IMF mass limit ($\msun$) for a user defined IMF, may be used multiple times, from \\
     & & & low to high mass\\
      -B & 0--$N/2$ & 0 & Number of binary systems (specify either $B$ or $b$)\\                        
      -b & 0.0--1.0 & 0.0 & Binary fraction (specify either $B$ or $b$)\\           
      -p & 0/1 & 1 & Binary pairing (random/ordered for $M > 5 \msun$)\\  
      -s & $s\ge 0$ & 0 & Seed for randomisation, set 0 for randomisation by local time\\   
      -t & 0/1/2/3 & 3 & Tidal field (no tidal field/near-field approximation/point-mass potential/\\
      & & & Milky-Way potential)\\           
      -e & $e\ge0$ & 0 & Epoch for stellar evolution in Myr (only available in \texttt{mcluster\_sse})\\                
      -Z & 0.0001-0.03 & 0.02 & Metallicity ($Z = 0.02$ for solar metallicity)\\          
      -X & $X\ge0$ & 8500/0/0 & Galactocentric radius vector in parsec (use 3 times for x-, y- and z-coordinate)\\       
      -V & $V\ge0$ & 0/220/0& Cluster velocity vector in km/s (use 3 times for x-, y- and z-coordinate)\\          
      -u  & 0/1 & 1 &Output units ($N$-body units/astrophysical units)\\           
      -h/-? & --- & --- & Display help                                        

\end{tabular}
\end{center}
\label{overviewtable}
\end{table*}

\section{McLuster manual}
The tool \textsc{McLuster} is an open source programme that can be used to either set up initial conditions for $N$-body computations or, alternatively, to generate artificial star clusters for direct investigation. There are two different versions of the code, one basic version for generating all kinds of unevolved clusters (in the following called \texttt{mcluster}) and one for setting up evolved stellar populations at a given age. The former is completely contained in the \textsc{C} file \texttt{main.c}. The latter (here dubbed as \texttt{mcluster\_sse}) is more complex and requires additional \textsc{FORTRAN} routines, namely the Single-Star Evolution (\textsc{SSE}) routines by \citet{Hurley00} which are provided with the \textsc{McLuster} code. For a quick introduction read the README file which is also provided with the code. For technical details on how to generate initial conditions for star cluster in general we would like to refer to \citet{Kroupa08} and referenced literature therein. 

\subsection{Compilation}
After extracting the archive which can be obtained from the given web address\footnote{\texttt{www.astro.uni-bonn.de/\~{}akuepper/mcluster/mcluster.html\\or  www.astro.uni-bonn.de/\~{}webaiub/english/downloads.php}}, the basic version \texttt{mcluster} can be compiled on a \textsc{UNIX} system from the command line with\\\\
\texttt{> cc -lm -o mcluster main.c\\\\}
where \texttt{cc} may be replaced by the \textsc{C-}compiler available on your computer. It can also be compiled by using the Makefile, i.e.\\\\
\texttt{> make mcluster\\\\}
For the more complex version \texttt{mcluster\_sse} the Makefile has to be used. Type\\\\
\texttt{> make mcluster\_sse\\\\}
after which the programme should compile, generating an executable named \texttt{mcluster\_sse}. 
\begin{itemize}
\item Note that, when using the Makefile, you may have to change the \textsc{C-} and/or \textsc{FORTRAN-}compiler entry as well as the path of your compiler. 
\item In case you want to apply any change to the code make sure that you first delete all object files by typing\\\\
\texttt{> make clean\\\\}
before re-compiling the code.
\end{itemize}

\subsection{Input}
There are two ways of choosing the desired cluster parameters. One is to set the parameters manually within \texttt{main.c}, within the upper part of the code right at the beginning of the \texttt{main} routine where all variables are declared. Note that, after changing  the value of a variable, you have to compile the code again. The more convenient way is therefore to pass arguments to the code at the time of execution via the command line (see Tab.~\ref{overviewtable} for an overview of available options). Type\\\\
\texttt{> mcluster -h\\\\}
or\\\\
\texttt{> mcluster\_sse -h\\\\}
respectively, to get a quick help on the available parameters and their usage. 
\begin{itemize}
\item In case you have not specified a certain parameter, the default value as set within the \texttt{main} routine is used.
\item Not all parameters can be set via the command line, some have to be changed within the code.
\item All command line arguments are the same for \texttt{mcluster} and \texttt{mcluster\_sse}, except for the age parameter (\texttt{-e}) which is mentioned in more detail below.
\end{itemize}
 
\subsection{Density profile}
\textsc{McLuster} can generate star clusters with various radial density profiles (option \texttt{-P}). In all cases the total mass of the cluster has to be chosen separately (options \texttt{-M} or \texttt{-N}). The available profiles are:
\begin{enumerate}
\item The simplest option is the analytical \citet{Plummer11} profile (option \texttt{-P0}) which is the simplest (two-parameter only) stationary solution to the collisionless Boltzmann equation. For this profile only the half-mass radius has to be specified additionally (option \texttt{-R}, in parsec). The Plummer profile is in principle infinitely extended but gets automatically truncated at the theoretical tidal radius of the cluster in case a tidal field has been specified (see below).
\item A more sophisticated set of models is given by the distribution function of \citet{King66}. For this profile (option \texttt{-P1}) the half-mass radius and the dimensionless value of $W_0$ which specifies the model concentration (option \texttt{-W} ranging from 2.0, i.e. low concentration, to 12.0, i.e. high concentration) has to be specified. The underlying routine for generating the density distribution from the distribution function is based on \textsc{king0} by Douglas C.~Heggie (e.g. \citealt{Heggie92}). Note that the final density distribution gets scaled to exactly match the desired half-mass radius; the radius at which the density becomes zero does not necessarily match the theoretical tidal radius in this case.
\item \citet{Subr08} give a density profile which can be chosen to be mass segregated. For this density profile (option \texttt{-P2}) the half-mass radius and an additional mass-segregation parameter (option \texttt{-S}, ranging from 0.0--0.99, but $S<0.5$ for reasonable models in virial equilibrium) have to be chosen. For \texttt{-S0.0} the {\v S}ubr profile is equal to a Plummer profile. \textsc{McLuster} uses a slightly modified version of the \textsc{Plumix} routine by Ladislav {\v S}ubr to set up this kind of profile.
\item Young clusters in the Large Magellanic Cloud were found to follow a two-dimensional density profile consisting of a core and a power-law tail without visible tidal truncation. \citet{Elson87} give a simple analytical formula for such profiles which can be deprojected with \textsc{McLuster} and used to set up 3d star cluster models (option \texttt{-P3}). Since those models are in principle infinitely extended, the so-called EFF models do not get scaled to a certain half-mass radius, but rather require specification of a cut-off radius to which the profile should extend (option \texttt{-c}, in parsec). The central density then gets calculated automatically using the specified cluster mass (see below). In addition, the radius of the two-dimensional core (option \texttt{-r}, in parsec) and the slope of the power-law part of the profile (option \texttt{-g}) have to be chosen. Values for \texttt{g} larger than 1.5 usually yield stable solutions, observational values lie at about 2.0 for young star clusters in the LMC \citep{Elson87}. The velocity field of this family of profiles is generated using the algorithm described in \citet{Kroupa08}.
\item A further possibility to set up the density distribution is given by option \texttt{-P-1}. In this case the final cluster has no density gradient, but consists of stars which are homogeneously distributed within a sphere. The size of the sphere is specified by choosing the half-mass radius (thus, the limiting radius of the sphere will be a factor $2^{1/3}$ larger). This option is especially useful in case of fractal initial conditions (see below). The velocities of the stars are isotropic and drawn from a Gaussian distribution. 
\end{enumerate}
\begin{itemize}
\item The exact matching of the actual half-mass radius as resulting from the discretized model to the specified value may be switched off within the \texttt{main} routine, but is not recommended. Set \texttt{match = 0} and compile the code again.
\end{itemize}

\subsection{Tidal field}
As mentioned above, the choice of the tidal field and of the cluster orbit may influence the extent of the density profile. \textsc{McLuster} offers different kinds of tidal fields which can be specified with the option \texttt{-t}. In addition it may be necessary to specify the galactocentric radius vector and the orbital velocity vector. This can be done by using the option \texttt{-X} (in parsec) and \texttt{-V} (in km/s), respectively, three times for the x-, y- and z-component (within a Cartesian coordinate system where the galactic disk would lie in the x-y-plane, if applicable). As an example, \texttt{-X8500.0 -X0.0 -X0.0 -V0.0 -V220.0 -V0.0} would give the motion of the Local Standard of Rest. This option is especially useful when generating input for $N$-body computations, since the input parameter file is automatically adjusted accordingly.
\begin{enumerate}
\item For a cluster in isolation choose \texttt{-t0}. No truncation is applied to profiles like the Plummer profile in this case.
\item A linearized tidal field, as described in \citet{Fukushige00} can be chosen with \texttt{-t1}. If you have selected to generate input files for \textsc{Nbody6} (see below) then the values for the Local Standard of Rest are used. In all other cases the galactocentric distance has to be specified additionally. Therefore use the option \texttt{-X} and set all but one component to zero, e.g. \texttt{-X6000.0 -X0.0 -X0.0} for an orbit at 6 kpc. No orbital velocity vector has to be specified as the linearized tidal field mimics a circular orbit.
\item If you choose \texttt{-t2} then you get a point-mass galaxy for which you can specify any galactocentric radius and orbital velocity (options \texttt{-X} and \texttt{-V}). The mass of the galaxy is set within the header of the \texttt{main} routine. By default, \texttt{M1pointmass} is set such that you get an orbital velocity of 220 km/s at a galactocentric radius of 8.5 kpc.
\item For a more realistic, Milky-Way potential you can choose option \texttt{-t3}. This potential consists of a central point mass/bulge, modelled as a Hernquist potential \citep{Hernquist90}, a Miyamoto disk \citep{Miyamoto75} and a logarithmic (phantom dark-matter) halo, with values as given in \citet{Allen91}. If you set up initial condition for \textsc{Nbody6} then the logarithmic halo will be adjusted such that the circular velocity, \texttt{VCIRC}, at some radius, \texttt{RCIRC}, has a specific value. The default is 220 km/s and 8.5 kpc, respectively, which may be changed in the header of the \texttt{main} routine. The other parameters of this potential may also be changed there.
\end{enumerate}

\subsection{Cluster mass and stellar mass function}
You can either fix the total number of stars in your cluster (option \texttt{-N}) or you can set a desired total mass (option \texttt{-M}, in solar units). In the latter case, \textsc{McLuster} draws stars from the selected mass function until the desired mass is exceeded. The mass function of stars in the cluster can be defined to be one of the following three kinds (option \texttt{-f}).
\begin{enumerate}
\item All stars can have the same mass (option \texttt{-f0}). The mass of each star is by default assumed to be $1\msun$, which may be changed within the \texttt{main} routine (parameter \texttt{single\_mass}).
\item The canonical \citet{Kroupa01a} initial mass function can be used with \texttt{-f1}. This IMF has a slope of $\alpha_1 = -1.3$ for stellar masses $m = 0.08 - 0.5 \msun$, and the Salpeter slope
$\alpha_2 = -2.3$ for $m > 0.5 \msun$. The lower and upper IMF limit, \texttt{mlow} and \texttt{mup}, are by default $0.08\msun$ and $100\msun$, respectively, but these values may be changed in the \texttt{main} routine. 
\item More sophisticated, multi-power-law IMFs can be set up with option \texttt{-f2}. Therefore, \textsc{McLuster} uses the routine \textsc{mufu} by Ladislav {\v S}ubr. This routine allows to define several mass limits and corresponding mass-function slopes between these limits. The limits and the slopes can be passed to \textsc{McLuster} with the options \texttt{-m} and \texttt{-a}, respectively. These options have to be used multiple times, where one more limit has to be passed to \textsc{McLuster} than number of slopes. For example, the Kroupa IMF with a steeper slope of $\alpha_3 = -2.7$ for stars more massive than $5\msun$ up to a maximum mass of $80\msun$ would be \texttt{-f2 -m0.08 -a-1.3 -m0.5 -a-2.3 -m5.0 -a-2.7 -m80.0}.
\end{enumerate}
\begin{itemize}
\item The  maximum stellar mass -- cluster mass relation found by \citet{Weidner06} can be used to automatically cut off the IMF at the corresponding upper mass limit. This routine is switched off by default but may be activated by setting the \texttt{weidner} parameter in the \texttt{main} routine to 1.
\item In \textsc{McLuster} there is a maximally allowed stellar mass limit defined through the parameter \texttt{upper\_IMF\_limit}. This parameter is set to $100\msun$ since \textsc{Nbody6}, i.e. the stellar evolution routine \textsc{SSE} within \textsc{Nbody6}, does not allow higher masses. In case you need higher stellar masses anyway, set this parameter within the \texttt{main} routine to the desired value but keep in mind that stars with mass larger than $100\msun$ are evolved as $100\msun$ stars.
\end{itemize}

\subsection{Mass segregation}
With \textsc{McLuster} it is possible to apply any degree of primordial mass segregation to all available density profiles, not only to the {\v S}ubr profile as already mentioned above. Therefore, the method as described in \citet{Baumgardt08a} is used. The advantage of this method is that the chosen density profile is not changed with increasing degree of mass segregation as is the case for the {\v S}ubr profile. 

In short, it works as follows: for a cluster of $N$ stars \textsc{McLuster} first draws the stellar masses from the selected IMF (see above) and then creates $N^\prime = N \langle m\rangle / m_{low}$ orbits, where $\langle m \rangle$ is the mean stellar mass and $m_{low}$ the lowest stellar mass in the cluster. These orbits get ordered by their specific energy, from low energy to high energy orbits. Then the stellar masses are also ordered and the cumulative mass function, $M_{cum}(i) = \sum_{j=1}^i M(j)$ is evaluated from this mass array, whereby $i=1\ldots N$. After dividing $M_{cum}(i)$ by the total cluster mass, the function is normalised such that it runs from 0 to 1. Finally, for any star an orbit from the list of energy-ordered orbits is chosen from the orbits between $N^\prime M_{cum}(i-1)$ and $N^\prime M_{cum}(i)$.

If the masses in the mass array are perfectly ordered from highest to lowest then this will yield a completely mass segregated cluster. That is, the highest mass star is on the lowest energy orbit, and the lowest mass star is on the highest energy orbit. 

Intermediate degrees of mass segregation can be achieved by non-perfect ordering. In \textsc{McLuster} this is realised as follows: first, all $N$ stellar masses are ordered from highest to lowest. Then, beginning with the highest mass, the masses are written to a new array, where the $i$-th massive star is written to the $j$-th empty slot counting from 0 to $N-i$. $j$ is generated using 
\begin{equation}
j = (N-i)\left(1-X^{1-S}\right),
\end{equation} 		
where $X$ is a random number between 0 and 1, and $S$ is the mass segregation parameter. When $S$ is zero, $j$ can have all values from 0 to $N-i$ and we end up with a random distribution. But when $S$ is 1, then $j$ is always zero and we reproduce the perfectly ordered array we started with. That is, because every star $i$, beginning with the most massive one, gets written to the next empty slot which is slot $i$. By choosing $S$ values between 0 and 1 we can get intermediate degrees of partial mass segregation (option \texttt{-S}). Unlike for the {\v S}ubr profile, $S$ can explicitly be chosen to be 1.0. Moreover, for all values of $S$ we get clusters in virial equilibrium (if not explicitly specified differently) with the desired (mass) density profile.

\subsection{Fractality}
Star clusters are not born in perfect symmetry. They are rather formed in collapsing, fractal molecular clouds (see e.g. \citealt{Gutermuth08, Koenyves10}). With \textsc{McLuster} you can set up two kinds of fractal initial conditions. First, you can set up a fractal distribution of stars within a sphere of constant average density, similarly as described in \citet{Goodwin04}. Secondly, you can add fractal substructure to any of the above given density profiles.
\begin{enumerate}
\item When you choose a homogeneous density profile (option \texttt{-P-1}) the cluster stars get distributed within a sphere as follows. The first star (a so-called parent) is placed in the middle of a box of size 2 (arbitrary units), then this box is split into 8 sub-boxes of half the initial box size. In the centre of each sub-box a further star is placed (a so-called child), whereupon each sub-box is split up into 8 smaller pieces, such that each child now becomes a parent on its own. By applying a small random offset to each star from its sub-box centre, we make sure that the final cluster does not look too grid-like. This is repeated until we have generated $128.0\times8^{\log(N)/\log(8)}$ stars or until the total number of stars would exceed this number with the next generation of children. From these stars we randomly draw $N$ stars with radial distances of less than unity from the centre of the initial box. We end up with a homogeneous sphere of stars.

Now, if not every sub-box gets a new star, and only those sub-boxes get sub-divided which have a star in their centre, then the final distribution of stars becomes fractal. The probability that a sub-box gets a star can be expressed as $2^{(D-3)}$, where $D$ is the fractal dimension (option \texttt{-D}). If $D$ is chosen to be 3.0 then we get no fractality since the probability is unity, i.e. every parent gets 8 children. If it is, e.g., 2.0 then only every second sub-box gets a star, or, on average, every parent has 4 children. 

Corresponding stellar velocities are drawn from a Gaussian distribution, and re-scaled such that all children of one parent are in virial equilibrium and the total mean velocity in one sub-group is unity (arbitrary units). In addition, each child gets the velocity of its parent. In a later step, \textsc{McLuster} re-scales the phase-space coordinates of the stars such that the cluster is in virial equilibrium (if not specified differently). In this way, we get a fractal structure consisting of coherently moving, gravitationally bound substructures. 

\item Alternatively, we can choose to set up fractal clusters which follow a given density profile like, e.g., the Plummer profile or the King profile, but which show fractal substructure. This is realised by first generating a sphere of stars of radius unity with the above procedure. But now this distribution of stars gets folded with the chosen density profile. Therefore the radius of each star first gets re-scaled by its absolute value to the power of three. In this way, we get $N$ stars with radii distributed homogeneously between 0 and 1, but which show substructure in 6D phase space. These radii are used as seeds to compute a corresponding radius within the specified density profile. In a last step, the space coordinates of each star get scaled by this newly generated radius. In this way the fractal distribution is conserved but folded with the specific density profile. Moreover, the velocity of each star is scaled to the expected velocity of a star at the given radius within the specified density profile. 

Note that the method described here is an ad hoc introduction of substructure which, like all other methods for generating substructured models, does not rely on any physical motivation. In this way, the generated clusters can have any degree of substructure and a smooth transition from spherical symmetry to substructuredness can be realised. This may be useful in some applications but is not meant to accurately reproduce observational substructure. In fact, due to the radial re-scaling the substructure gets stretched which produces long filaments in the final clusters. These filaments may be compared to the filamentary structure of molecular gas in star forming regions.
\end{enumerate}
\begin{itemize}
\item In order to guarantee a minimum of spherical symmetry, you can tell \textsc{McLuster} to give 8 children to the first ``ur-parent''. In this way, you avoid too large asymmetries. This will lead to less differing results between initial conditions generated with different random seeds, but does on the other hand not yield perfectly fractal clusters. This tweak can be switched off within the header of the \texttt{main} routine. Set the \texttt{symmetry} parameter to 0 and re-compile. 
\item Once in a while \textsc{McLuster} gets stuck in the fractality sub-routine. In this case a restart with a different random seed should help.
\end{itemize}

\subsection{Binaries}
After the stellar masses got drawn from an initial mass function, you can choose to let \textsc{McLuster} set up binary systems. You can specify either the desired number of binary systems (option \texttt{-B}, values from 0 to $N$/2) or you can specify a fraction of stars which should be in binary systems (option \texttt{-b}, values from 0 to 1). Thus, from all $N$ stars, $N\times b/2$ or $B$ binaries are formed, respectively. The binaries are then replaced by a centre-of-mass (CoM) particle for the rest of the procedure. Only in the very end, after the density profile has been established and the velocities of the cluster members have been scaled appropriately, the CoM particles get replaced by their two constituent stars. The binary orbital plane is oriented randomly and the orbital phase is also chosen randomly. The internal properties of the binaries can be generated according to the following semi-major axis distributions which can be selected in the header of the \texttt{main} routine (parameter \texttt{adis}). 
\begin{enumerate}
\item A flat distribution of semi-major axes can be specified with \texttt{adis = 0}. In addition, you have to choose a minimum and a maximum semi-major axis (parameters \texttt{amin} and \texttt{amax}).
\item If \texttt{adis } is set to 1 (default) then the semi-major axis of each binary is computed from a period which was drawn from the \citet{Kroupa95a} period distribution (see also \citealt{Kroupa08}). This is the preferred distribution function as it unifies the observed Galactic-field and the pre-main-sequence population.
\item If you want the semi-major axes to be generated from the \citet{Duquennoy91} Galactic-field period distribution then you have to set \texttt{adis = 2}.
\end{enumerate}
\begin{itemize}
\item The pairing of primary and secondary components of the binaries can be chosen to be either random or ordered. In the latter case (option \texttt{-p1}) the stellar masses above a certain threshold (parameter \texttt{msort} in the \texttt{main} routine) get ordered from most to least massive. The rest of the stars are put in random order onto the list below the last star with mass above \texttt{msort}. The pairing of binaries now starts with the most massive star which gets paired together with the second-most massive, then the third with the forth, and so on down the list. Below \texttt{msort} pairing is random, which is consistent with the binary-star observational data \citep{Kroupa08}. The default value of \texttt{msort} is $5\msun$, in rough agreement with recent findings (e.g. \citealt{Kobulnicky07}). 
\item Eccentricities, $e$, are drawn from a thermal eccentricity distribution, i.e. $f(e) = 2e$ (e.g. \citealt{Duquennoy91}; see also \citealt{Kroupa08}).
\item To correct for the fact that short-period binaries in the Milky Way do not show high eccentricities \citep{Mathieu94}, \citet{Kroupa95b} introduces an analytical correction for such systems, which is attributed to pre-main sequence eigenevolution between the constituent stars. This correction can be switched off by setting the parameter \texttt{eigen = 0} in the \texttt{main} routine.
\end{itemize}
A discussion of binary systems and their formation can be found in \citet{Kroupa09}.
\subsection{Stellar evolution}
\textsc{McLuster} contains the \textsc{SSE} routines by \citet{Hurley00} which are also used in, e.g., \textsc{Nbody6} for stellar evolution. If you just want to generate star clusters consisting of zero-age main sequence (ZAMS) stars then you only need the basic version \texttt{mcluster}. But if you want to set up a cluster with an evolved stellar population then you have to use \texttt{mcluster\_sse} which makes use of those \textsc{SSE} routines. In this case you have to specify an age for the cluster population (option \texttt{-e}, in Myr). The evolution of the stars is done in the very beginning of the programme. When the stars are drawn from the IMF they get immediately evolved to the desired age. The masses of the evolved stars are then summed up and additional stars are generated until the desired cluster mass is exceeded or the desired number of stars is reached. The stellar parameters derived from \textsc{SSE} for each star are stored in an additional file, which also has to be passed to \textsc{Nbody6} (see below).
\begin{itemize}
\item The internal parameters of \textsc{SSE} can be changed within the header of the \texttt{main} routine (not recommended).
\item In case a star becomes a neutron star or a black hole \textsc{SSE} assigns a kick velocity to the remnant (if not specified differently). The kick velocity can be used to remove the remnant from the cluster. Therefore the present-day escape velocity of the cluster at its half-mass radius is calculated and if the kick velocity exceeds this velocity it gets removed. If you want to keep all compact remnants set the parameter \texttt{remnant = 0} in the \texttt{main} routine.
 \item The metallicity, $Z$, can be set with option \texttt{-Z}. Alternatively, you can specify the metallicity as [Fe/H] within the \texttt{main} routine, the corresponding $Z$ value is computed using the relation given in \citet{Bertelli94}. Make sure that in this case you set $Z = 0$ beforehand. 
\item In case you are generating binaries and have selected ordered pairing for stars above a certain mass, \texttt{msort} (see above), then the ZAMS mass is used to decide whether a star is paired randomly to another star or not.
\item The components of binaries are independently evolved as single stars with \textsc{SSE}. For a more advanced treatment of stars in binaries, the Binary-Star Evolution (\textsc{BSE}) routines by \citet{Hurley02} are also included in \textsc{McLuster}. Therefore, at the very end of the cluster generation procedure, when the binary CoM particles are replaced by their constituents and the orbital elements are generated, the masses of the components are reset to their ZAMS mass. Then the two stellar masses, a semi-major axis and an eccentricity are passed to \textsc{BSE} which finally returns corresponding evolved values. This feature is switched off by default but may be activated by setting the parameter \texttt{BSE} to 1 in the \texttt{main} routine.
\end{itemize}

\subsection{Output}
Up to now \textsc{McLuster} can generate input for \textsc{Nbody6} (option \texttt{-C0}, \citealt{Aarseth03}) and \textsc{Nbody4} (option \texttt{-C1}), or it can write an ASCII table of stars and their properties (option \texttt{-C3}). 
\begin{enumerate}
\item In the first and second case, there will be two output files which can be named with option \texttt{-o}. For example, \texttt{-o mycluster} will yield the files \texttt{mycluster.input}, containing all the input parameters for the run, and \texttt{mycluster.fort.10}, containing the masses, positions and velocities. Note that the latter has to be renamed to \texttt{fort.10} at the time of execution in order to be recognised by \textsc{Nbody4/6}. When using \texttt{mcluster\_sse} there will be another file named \texttt{mycluster.fort.12}. This file also has to be renamed within the directory of the run to \texttt{fort.12}. The name \texttt{mycluster} is just added to the file names for convenience. Thus, a directory for an  \textsc{Nbody4/6} run should contain:
\begin{enumerate}
\item \texttt{mycluster.input},
\item \texttt{fort.10},
\item \texttt{fort.12}.
\end{enumerate}
The run is then started with the usual command, i.e., \\\\
\texttt{> /.../nbody6 < mycluster.input}\\\\
where \texttt{/.../} should be replaced by the path to your \textsc{Nbody4/6} installation. 
\item In the case of \texttt{-C3} a file named \texttt{mycluster.txt} will be created containing mass, positions ($x$, $y$, $z$) and velocities ($v_x$, $v_y$, $v_z$). If you are using \texttt{mcluster\_sse} then this file will also contain for each star the ZAMS mass\footnote{from \textsc{SSE}, in $\msun$}, the stellar type\footnote{from \textsc{SSE}, see \citealt{Hurley00}}, the epoch of the star\footnote{from \textsc{SSE}, in Myr}, its spin\footnote{from \textsc{SSE}, in km/s}, its radius\footnote{from \textsc{SSE}, in solar units}, its luminosity\footnote{from \textsc{SSE}, in solar units}, its age  in Myr, its metallicity ($Z$), its absolute $V$ magnitude, its apparent $V$ magnitude\footnote{assuming a distance \texttt{Rgal} from the observer which can be changed in the \texttt{main} routine}, $B-V$, its effective temperature (K), a random error for the $V$ magnitude, and a random error for $B-V$. The last six are generated assuming observations with an 8m-class telescope from a distance \texttt{Rgal} (see \citealt{Kuepper11}). 
\end{enumerate}
\begin{itemize}
\item In addition you have to specify whether you want the output to be in $N$-body units (see e.g. \citealt{Heggie03}, option \texttt{-u0}) or astrophysical units (option \texttt{-u1}). For the \textsc{McLuster} output to serve as input for \textsc{Nbody6} and \textsc{Nbody4} this output should always be in $N$-body units.
\item With the ASCII table output you can easily draw a colour-magnitude diagram. Use columns 18(+21) versus 17(+20) for a diagram showing $B-V$ versus apparent $V$ magnitude (+random errors).
\item \textsc{McLuster} automatically computes a radial density profile and a cumulative radial density profile. Both are by default printed to the screen. This may be switched off within the \texttt{main} routine (parameters \texttt{create\_radial\_profile = 0} or \texttt{create\_cumulative\_profile = 0}, respectively).
\end{itemize}

\subsection{Miscellaneous}
\begin{itemize}
\item The virial ratio, $Q = -E_{kin}/E_{pot}$, where $E_{kin}$ is the total kinetic energy of the single cluster stars and $E_{pot}$ their potential energy, can be set with the parameter \texttt{-Q}. Note that this only affects the input file for $N$-body computations but not the stellar velocities in the table of stars (there the virial ratio will always be 0.5, i.e. virial equilibrium). The velocities get scaled within \textsc{Nbody4/6} according to your choice of $Q$. 
\item The random seed can be set with the parameter \texttt{-s} which can be any positive integer. In the case of \texttt{-s0} \textsc{McLuster} will take the local time as random seed.
\item There is an upper limit of temporary stars or orbits within \textsc{McLuster}. This number, \texttt{Nmax}, is set to 1.500.000 by default, i.e. \textsc{McLuster} allocates memory accordingly. When applying mass segregation or fractality to a cluster, many more temporary stars/orbits have to be generated than finally needed. Especially if a cluster shall be mass segregated and fractal at the same time this number may easily be exceeded. In this case you should increase it in the \texttt{main} routine. 
\item A few more parameters and command line arguments are available (see option \texttt{-h} and the header of the \texttt{main} routine) which mostly affect flags for $N$-body computations.
\end{itemize}

\subsection{Examples}
If no command line arguments are passed to \textsc{McLuster} it will use the default parameter values which are specified and which can be changed within the \texttt{main} routine. By typing\\\\
\texttt{> mcluster}\\\\
a file \texttt{test.txt} is created. This default cluster has $1000\msun$ ($\sim 1800$ stars), a half-mass radius of 0.8 pc, a Plummer density distribution, is in a Milky-Way tidal field with LSR values, uses the Kroupa IMF and has no binaries. The entries in the data table are in astrophysical units. With\\\\
\texttt{> mcluster -C0 -u0}\\\\
the same cluster is written to the files \texttt{test.input} and \texttt{test.fort.10} but in $N$-body units. This can be passed to \textsc{Nbody4/6} as stated above. The clusters used in this work were created using, e.g.,\\\\
\texttt{> mcluster -M100000.0 -P3 -r0.1 -c20.0 -g2.0 -S1.0 -C0 -G1 -o R136 -f1 -b0.2 -p1 -s2 -Z0.01 -u0}\\\\
for the fully mass segregated $N$-body model. The arguments stand for: a total mass of $100.000\msun$ (\texttt{-M}), the EFF density profile (\texttt{-P}) with a 2d core radius of 0.1 pc (\texttt{-r}), a cut-off radius of 20 pc (\texttt{-c}) and a 2d power-law slope of -2 (\texttt{-g}). It is completely mass segregated (\texttt{-S}), the output is for \textsc{Nbody6} (\texttt{-C}), and we use a GPU (\texttt{-G}). The output is named R136 (\texttt{-o}), we use a Kroupa IMF (\texttt{-f}), 20\% binaries (\texttt{-b}) and ordered pairing for massive stars (\texttt{-p}). The random seed of our model is 2 (\texttt{-s}) and the metallicity is 0.01 (\texttt{-Z}). The output is in $N$-body units (\texttt{-u}) since we want to pass it to \textsc{Nbody6}.